\documentclass[aps,prl,10pt,amsmath,superscriptaddress,twocolumn,floatfix,longbibliography]{revtex4-2}
\usepackage[T1]{fontenc}
\usepackage[utf8]{inputenc}
\usepackage{times,graphicx,amsmath,amssymb,booktabs,bm,xcolor}
\usepackage[colorlinks=true,citecolor=blue,linkcolor=purple,urlcolor=blue]{hyperref}
\newcommand{\ket}[1]{\lvert #1\rangle}
\newcommand{\bra}[1]{\langle #1\rvert}
\newcommand{\Tr}{\operatorname{Tr}}
\newcommand{\dket}[1]{\lvert #1\rangle\!\rangle}
\newcommand{\dbra}[1]{\langle\!\langle #1\rvert}
\begin{document}
\title{Distinguishing Conserved Quantum Sectors via Measurement Record Features}
\author{Hao Zhu}
\affiliation{State Key Laboratory of Semiconductor Physics and Chip Technologies, Institute of Semiconductors, Chinese Academy of Sciences, Beijing 100083, China}
\author{Guo-Feng Zhang}
\email{gf1978zhang@buaa.edu.cn}
\affiliation{School of Physics, Beihang University, Beijing 100191, China}
\author{Wen-Long Ma}
\email{wenlongma@semi.ac.cn}
\affiliation{State Key Laboratory of Semiconductor Physics and Chip Technologies, Institute of Semiconductors, Chinese Academy of Sciences, Beijing 100083, China}
\affiliation{Center of Materials Science and Opto-Electronic Technology, University of Chinese Academy of Sciences, Beijing 100049, China}
\date{\today}

\begin{abstract}
Conservation laws partition the Hilbert space of a quantum system into invariant sectors that are often difficult to identify directly. Repeated measurements on an ancillary probe, however, generate a measurement record that provides information about the occupied sector. Here we develop a framework for identifying conserved sectors using a $(K+1)$-dimensional feature vector of record statistics, comprising the average outcome frequency and pair-outcome correlations up to $K$ measurement cycles apart. For fixed $K$, the feature vector can be evaluated at a cost linear in record length. In quantum non-demolition measurement models, the average outcome frequency suffices for accurate sector identification. For an interacting spin bath with nontrivial dynamics within each sector, incorporating temporal correlations between successive and more widely separated measurements into the record feature vector substantially improves classification fidelity. Our framework provides an efficient approach to conserved-sector identification and demonstrates the importance of temporal correlations in distinguishing the underlying sector dynamics.
\end{abstract}
\maketitle

\textit{Introduction.---}
Conservation laws partition the Hilbert space of a quantum system into invariant sectors~\cite{buca2012,Albert2014,sanchezMunoz2019}, which are often labeled by conserved charges, excitation numbers, or generalized parities. Such sectors are useful for symmetry-protected quantum memories~\cite{lidar1998}, conditional state preparation~\cite{Cirac1993,dasari2022purification}, and quantum error detection~\cite{Jin2025,Hung2026}. Identifying the occupied sector in an individual experiment can be difficult when the corresponding conserved operator cannot be measured directly.
Instead, indirect readout, often mediated by a coupled ancillary probe, can provide access to sector labels across platforms including cavity and circuit QED~\cite{Girvin2014,Blais2021,guerlin2007photon}, solid-state spins~\cite{Burkard2023,Friesen2004,HarveyCollard2018,CirianoTejel2021,Oakes2023,Gachter2022,Nurizzo2023,DAnjou2019,Ginzel2023,Serrano2024,Mielke2021,Debnath2021,Rosenthal2024,xue2020soft}, and atomic systems~\cite{Huie2023,Lee2026,Hilder2022},
with applications in quantum-state preparation and purification~\cite{Cirac1993,dasari2022purification}, readout and sensing~\cite{guerlin2007photon,hann2018robust,elder2020highfidelity,sun2014parity,Hilder2022,marcks2024charge}, and quantum error detection~\cite{Jin2025,bultink2020leakage,Hung2026}.
% with applications in photon counting~\cite{guerlin2007photon}, Fock-state preparation~\cite{Cirac1993}, bosonic and multilevel readout~\cite{hann2018robust,elder2020highfidelity}, spin-bath purification~\cite{dasari2022purification}, parity detection~\cite{sun2014parity,Jin2025,Hilder2022}, leakage detection~\cite{bultink2020leakage,Hung2026}, and charge sensing~\cite{marcks2024charge}.

Repeated measurements of an ancillary probe~\cite{guerlin2007photon,xue2020soft,Cech2023} generate temporal records that provide information about the quantum system~\cite{Brun1997,wiseman1996}.
These records have been used for parameter estimation~\cite{gammelmark2013,kiilerich2016} and quantum-trajectory reconstruction~\cite{flurin2020,koolstra2022}, while the statistics of quantum trajectories have been studied to characterize fluctuations~\cite{garrahan2010,lesanovsky2013,carollo2019,Flindt2013,Znidaric2014,Paulino2024,Li2026,Cech2023,Kleinherbers2022} and dynamical phenomena in monitored quantum systems~\cite{Vijay2011,Lee2012,Chen2021,Ashida2018,ferrari2025dissipative,mochizuki2025spectral,li2025monitored,cea2026large,ma2018phase,Cech2026}.
In ideal quantum nondemolition (QND) readout~\cite{jordan2005kicked,ralph2006qnd,wang2023synthesize}, the average outcome frequencies can reliably distinguish conserved sectors without internal dynamics~\cite{bauer2011,bauer2013,ma2023}.
However, nontrivial internal sector dynamics can affect the statistics of probe measurement records, allowing distinct sectors to exhibit identical outcome frequencies but different temporal correlations.
These temporal correlations can provide substantially more information about the underlying sector dynamics than average outcome frequencies alone.
Recent studies have shown that correlations between measurement outcomes at different steps can be used to learn quantum environments~\cite{liu2010correlations,wang2019arbitrary,pfender2019} and many-body dynamics~\cite{cheung2024faraday,wu2024selective}, with related methods applied to quantum sensing~\cite{shen2023quantum,montenegro2022sequential,Ilias2025} and system learning~\cite{zhan2024conservation,dong2026}.
Statistical inference~\cite{tsang2012,DAnjou2014a,DAnjou2016,barratt2022learnability}, hidden Markov models~\cite{Martinez2020,Spethmann2025,bultink2020leakage}, machine-learning methods~\cite{Liu2020,angelatos2021,koolstra2022,dehghani2023}, and soft-information decoding~\cite{DAnjou2014b,Ali2024,Berent2024,Raveendran2023,Majaniemi2026} have been used to improve the extraction of system information from measurement records.
However, it remains unclear how to systematically select and combine a compact set of temporal correlations for efficient conserved-sector identification.

\begin{figure}[tbp]
\includegraphics[width=\linewidth]{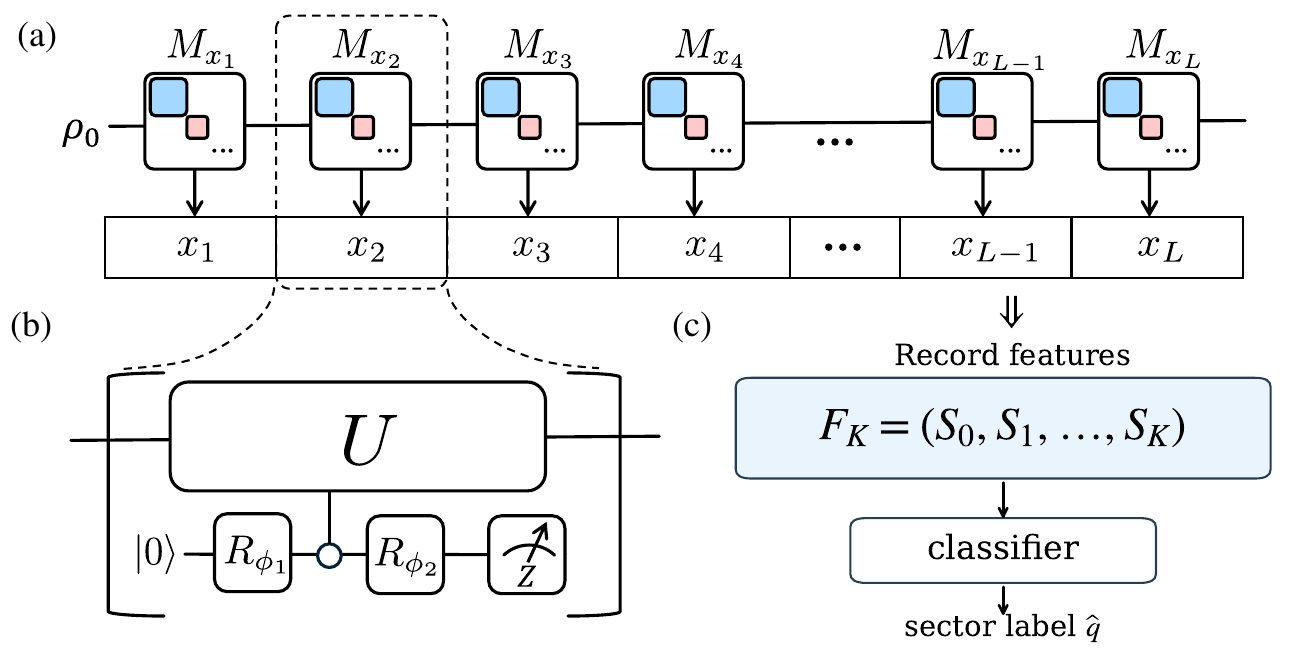}
\caption{Sector identification via repeated-measurement record features. (a) Repeated probe measurements generate a length-$L$ measurement record $(x_1,\cdots,x_L)$ with $x_n=\pm 1$. (b) Each measurement can be implemented by a RIM cycle that resets and measures the probe, and the measured system retains its state between adjacent RIMs. (c) The sector classifier is based on the record feature $F_K=(S_0,\ldots,S_K)$ with $K\ll L$, where $S_0=\frac{1}{L}\sum_{n=1}^Lx_n$ is the outcome sum, and $S_i=\frac{1}{L-i}\sum_{n=1}^{L-i}x_nx_{n+i}$ is the pair sum at lag $i\ge1$.}
\label{fig:protocol}
\end{figure}

In this Letter, we develop a framework for efficiently identifying conserved quantum sectors using a multidimensional feature vector constructed from probe measurement records (Fig.~\ref{fig:protocol}). The record feature $F_K$ combines the mean outcome with pair correlations up to an adjustable lag $K$ and can be evaluated at a cost linear in record length for fixed $K$. Such a record feature can capture sector information that is absent from single-outcome statistics. Classifiers trained on sector-labeled records are evaluated using independent test data. In QND models, simple average frequency statistics suffice for sector identification, whereas correlations between successive and more widely separated outcomes significantly improve classification in an interacting spin bath. We compare feature-based classifiers with a neural network (NN) using complete records and examine the benefits of including additional correlations. Our framework provides a systematic and efficient approach to conserved-sector identification by exploiting temporal correlations that distinguish the underlying sector dynamics.

\textit{Conserved quantum sectors.---}
An indirect measurement of a quantum system $S$ is described by outcome-resolved Kraus operators $\{M_x\}$, satisfying $\sum_xM_x^\dagger M_x=I_S$ with $I_S$ being the identity operator on the system, with the associated quantum channel $\mathcal E(\rho)=\sum_xM_x\rho M_x^\dagger$~\cite{wiseman1996,albert2019,ivashkov2024dynamic}.
Consider an orthogonal decomposition $\mathcal H_S=\bigoplus_q\mathcal H_q$, with projectors $\{\Pi_q\}$ satisfying $\sum_q\Pi_q=I_S$.
We call these subspaces conserved sectors when every measurement outcome preserves the decomposition,
\begin{align}
[M_x,\Pi_q]=0,\qquad \forall x,q.
\end{align}
For a trace-preserving channel, this condition is equivalent to $\mathcal E^\dagger(\Pi_q)=\Pi_q$, where $\mathcal E^\dagger(O)=\sum_xM_x^\dagger O M_x$ is the Heisenberg-picture adjoint~\cite{albert2019}.
Thus, the sector projectors are conserved observables, and no measurement outcome can induce transitions between different sectors.
Equivalently, all off-diagonal blocks $\Pi_rM_x\Pi_q$ vanish for $r\neq q$, giving $M_x=\bigoplus_qM_x^{(q)}$.
This decomposition prevents mixing between sectors while allowing nontrivial, potentially noncommuting dynamics within each sector.

To understand the microscopic origin of this structure, consider a target system $S$ coupled to an ancilla $a$ through
\begin{align}
H_{aS}=H_a\otimes I_S+I_a\otimes H_S+\sum_\nu A_\nu\otimes B_\nu,
\end{align}
where $H_a$ and $H_S$ are the free Hamiltonians, while $A_\nu$ and $B_\nu$ are the probe and system coupling operators, respectively. If the probe and system are initially uncorrelated, the reduced system evolution has the Stinespring form $\mathcal E(\rho)=\Tr_a[U(T)(\rho_a\otimes\rho)U^\dagger(T)]$, where $U(T)=e^{-iH_{aS}T}$ and $\Tr_a$ traces out the probe. For a pure initial probe state $\rho_a=\ket{\psi}_a\bra{\psi}$ and a projective readout in the basis $\{\ket{x}_a\}$, the corresponding Kraus operators are $M_x={}_a\langle x|U(T)|\psi\rangle_a$.
If $[H_S,\Pi_q]=[B_\nu,\Pi_q]=0$ for all $q,\nu$, then $[H_{aS},I_a\otimes\Pi_q]=0$, so the joint evolution preserves each $\mathcal H_a\otimes\mathcal H_q$. Each Kraus operator inherits the same commutation relations, $[M_x,\Pi_q]=0$, and hence the block-diagonal structure established above.

A typical mechanism is a conserved charge $Q$, whose eigenspaces remain isolated when $[H_S,Q]=[B_\nu,Q]=0$~\cite{buca2012,Albert2014,sanchezMunoz2019}.
More subtle invariant structures can arise through destructive interference, where coherent superpositions form dark subspaces decoupled from certain transitions~\cite{lidar1998,villazon2020}.
If the remaining dynamics also preserve these subspaces, their projectors define additional conserved sectors.
Thus, both symmetry-protected and dynamically generated sectors are characterized by the same Kraus-invariant decomposition.

A concrete realization is provided by Ramsey interferometric measurements (RIMs) of an ancillary qubit [Fig.~\ref{fig:protocol}(b)]. The probe interacts with the system through
\begin{align}
H_{aS}=I_a\otimes H_S+\sigma_a^z\otimes B.
\end{align}
We define a $\pi/2$ rotation of the probe as $R_\theta=\exp[-i\pi(\cos\theta\,\sigma_a^x+\sin\theta\,\sigma_a^y)/4]$. Each RIM cycle starts with the probe initialized in $\ket{0}_a$, followed by a rotation $R_{\phi_1}$ that prepares an equal superposition of its two $\sigma_a^z$ eigenstates. The probe and system then evolve jointly for a time $T$, during which the two probe states induce conditional system evolutions $U_\pm=\exp[-i(H_S\pm B)T]$. A second rotation $R_{\phi_2}$ is then applied, followed by a projective measurement of the probe in the $\sigma_a^z$ basis, with the outcome $\alpha\in\{0,1\}$ denoted by $x_n=(-1)^\alpha$. The resulting Kraus operators acting on the system are $M_{x_n}=\frac12(U_+-x_n e^{i\phi}U_-)$, where $\phi=\phi_1-\phi_2$ is the relative pulse phase, and outcome-dependent overall phases have been omitted. Since $M_{x_n}$ are linear combinations of $U_\pm$, any common invariant decomposition of $H_S$ and $B$ is preserved by the measurement. Repeating this cycle generates the temporal measurement records used to identify conserved sectors.

\textit{Repeated measurements.---}
A single probe outcome can be compatible with several conserved sectors. Repeated measurements allow sector-dependent information to accumulate in a record, including through temporal correlations between outcomes. A sequence of $L$ applications of the same protocol [Fig.~\ref{fig:protocol}(a)] produces a record $\bm x=(x_1,\ldots,x_L)$ with probability $P^{(L)}(\bm x)=\Tr(M_{\bm x}\rho_0M_{\bm x}^\dagger)$, where $M_{\bm x}=M_{x_L}\cdots M_{x_1}$. The probe is reset after each cycle, while the target retains the state conditioned on the preceding outcomes. This state influences later outcomes and can generate temporal correlations. Such conditioned evolution is described by quantum trajectories~\cite{wiseman1996,brun2002}.

For an initial state $\rho_0$, sector $q$ has weight $\mu_q=\Tr(\Pi_q\rho_0)$ and normalized state $\eta_q=\Pi_q\rho_0\Pi_q/\mu_q$ for $\mu_q>0$. Starting from $\eta_q$ defines the sector-conditioned record distribution $P_q^{(L)}(\bm x)$. Since every measurement preserves the sectors, the full record probability is
\begin{equation}
P^{(L)}(\bm x)=\sum_q\mu_qP_q^{(L)}(\bm x).
\label{eq:mixture}
\end{equation}
This relation also holds in the presence of initial coherences between sectors, since off-diagonal sector blocks have zero trace. The record statistics thus admit an equivalent two-step description: select sector $q$ with probability $\mu_q$, then generate a record according to $P_q^{(L)}(\bm x)$. Given an observed record of finite length, our task is to identify the underlying conserved sector.

\textit{Classifier via record feature vectors.---}
To construct a classifier, we first choose record statistics $F(\bm x)$ and collect training records of length $L$ from each known sector $q$. These records provide an estimate $\widehat p_{q,F}^{(L)}(f)$ of the feature distribution in that sector. For count statistics, a normalized histogram gives the fraction of training records with each value of $F$. Once these distributions are fixed, the classifier evaluates $F(\bm x)$ for a new record and obtains its estimated probability in each sector. Each probability is weighted by the initial sector weight $\mu_q$. The record is assigned to the sector with the largest weighted probability:
\begin{equation}
\widehat Q_F(\bm x)=\arg\max_q\left[\mu_q\widehat p_{q,F}^{(L)}(F(\bm x))\right].
\label{eq:classifier}
\end{equation}

For a measurement record with binary outcomes $x_n=\pm1$, we construct a $(K+1)$-dimensional record feature vector as
\begin{equation}
F_K=(S_0,S_1,\ldots,S_K).
\label{eq:statistics}
\end{equation}
Here $S_0=L^{-1}\sum_{n=1}^Lx_n$ measures the mean outcome, while $S_k=(L-k)^{-1}\sum_{n=1}^{L-k}x_nx_{n+k}$ measures the correlations between measurement outcomes at lag $k\ge1$ [Fig.~\ref{fig:protocol}(c)]. The corresponding counts are $N_+=L(1+S_0)/2$ for positive outcomes and $N_k=(L-k)(1+S_k)/2$ for pairs of identical outcomes at lag $k$. Thus $F_0$ uses only outcome counts, $F_1$ adds adjacent pairs, and $F_K$ retains pairs up to lag $K$. 
We extend the feature to longer lags because correlations at different lags can reveal different aspects of the dynamics within each sector.

For stationary, ergodic measurement records within a sector, the statistical average of $S_0$ converges to the stationary outcome mean as $L$ increases, while that of $S_k$ at fixed $k$ converges to the corresponding stationary correlation, i.e., $\mathbb E_q[S_0]=m_q$, $\mathbb E_q[x_nx_{n+k}]=C_q(k)$. If these limits form distinct vectors for different sectors, sufficiently long records concentrate around different feature values, enabling reliable sector identification. Explicit expressions for these centers, their convergence conditions, and finite-record fluctuations are given in the End Matter.

An NN classifier provides a benchmark with the cumulative-count input used in Ref.~\cite{Liu2020}. Its input is $z_\ell=\sum_{n\le\ell}(x_n+1)/2$ for every $\ell=1,\ldots,L$. Successive differences recover each outcome, so this input retains the full record. The NN has a hidden layer of 32 units and outputs a normalized score $s_{q,L}(\bm x)$ for each sector. We train with equal numbers of records per sector and select checkpoints on validation data. For a new record, the NN predicts $\widehat Q_{\rm NN}(\bm x)=\arg\max_q[\mu_qs_{q,L}(\bm x)]$.
Compared with the length-$L$ NN input, $F_K$ contains only $K+1$ averages, independent of $L$ at fixed $K$, and is inexpensive to evaluate, with cost linear in $L$. By supplying the temporal correlations explicitly, the feature-based classifier reduces the burden of learning informative patterns from full records and can capture sector information that a finitely trained NN fails to exploit. 

%As the examples below show, these compact features achieve accurate sector identification in the simple models and outperform the NN benchmark in the interacting bath.

To compare the classifiers on the same basis, we evaluate their predictions on the same independent test records. We define the \emph{classification fidelity} as the probability of correct sector identification, averaged over the initial sector weights:
\begin{equation}
\mathcal F_C(L)=\sum_q\mu_q P_q[\widehat Q_C(\bm x)=q],
\label{eq:fidelity}
\end{equation}
where $C$ denotes a feature-based or NN classifier. The performance panels plot the \emph{classification infidelity} $1-\mathcal F_C(L)$, the probability of an incorrect assignment. Comparing these errors reveals the gain from adjacent pairs, longer lags, or learned processing. We now apply this comparison to three physical settings with increasingly rich dynamics within each sector, from QND readout to an interacting spin bath. Together, they illustrate when outcome counts suffice and when temporal correlations provide additional sector information.

\begin{figure}[tbp]
\includegraphics[width=\linewidth]{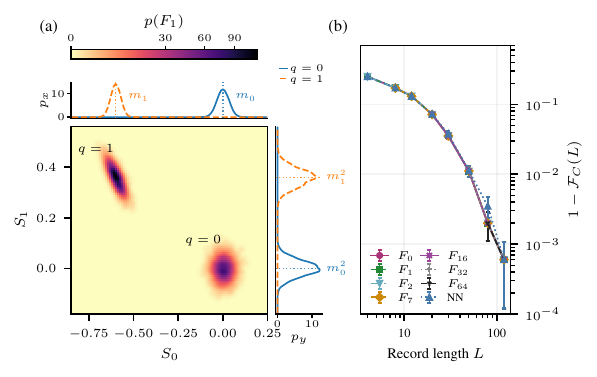}
\caption{QND readout with equal sector weights, $\phi=0$, $b_0T=\pi/4$, and $b_1T=\arccos\sqrt{0.8}$. (a) Joint density $p(F_1)=p(S_0,S_1)$ at $L=1000$, estimated from independent test records, with separately normalized sector-conditioned marginal densities above and to the right. Dotted guides mark $m_q$ and $m_q^2$ in the $S_0$ and $S_1$ marginals, respectively. The horizontal colorbar indicates the joint density. (b) Classification infidelities of the $F_K$ classifiers with $K=0,1,2,7,16,32,64$ and the full-record NN. All tested $F_K$ curves coincide. Error bars indicate approximate 95\% confidence intervals from finite test sampling.}
\label{fig:qnd}
\end{figure}

\textit{Example I: QND readout.---} Consider commuting normal Kraus operators, $M_x\ket{j}=c_{xj}\ket{j}$, with $|c_{xj}|^2$ independent of $j$ within each sector~\cite{ma2023}. The outcomes within sector $q$ are independent and identically distributed with mean $m_q$, giving $\mathbb E_q[S_0]=m_q$ and $\mathbb E_q[S_k]=m_q^2$ for $1\le k<L$. Their respective variances are $(1-m_q^2)/L$ and $(1+2m_q^2-3m_q^4)/L+O(L^{-2})$, both decaying as $L^{-1}$ for fixed $k$. Full finite-record distributions are given in the Supplemental Material~\cite{supplementaryMaterial}. The record likelihood depends only on $S_0$, making it a sufficient statistic at every record length. To illustrate these statistics, we consider the RIM realization with $H_{aS}=\sigma_a^z\otimes B$ and $B=\sum_{q=0,1}b_q\Pi_q$, giving $m_q=-\cos(2b_qT+\phi)$~\cite{supplementaryMaterial}. With equal priors, the $S_0$ and $S_1$ distributions concentrate around distinct centers as records grow longer [Fig.~\ref{fig:qnd}(a)]. All tested $F_K$ classifiers achieve the same classification infidelity as the count-based classifier [Fig.~\ref{fig:qnd}(b)], while the NN approaches this optimum.

\textit{Example II: Spin-Pair Parity Readout.---}
We next consider parity sectors with nearly indistinguishable $S_0$ distributions but distinct $S_1$ statistics. The two dynamical sectors $q=\pm$ are described by
\begin{equation}
H_q=w_{xq}\sigma_x+w_{zq}\sigma_z,\qquad B_q=\kappa_q\sigma_z.
\label{eq:effectiveMarkov}
\end{equation}
Here $q=+$ and $q=-$ denote the even and odd subspaces, with bases $\{\ket{\uparrow\uparrow},\ket{\downarrow\downarrow}\}$ and $\{\ket{\uparrow\downarrow},\ket{\downarrow\uparrow}\}$, respectively.
These blocks arise from an anisotropically coupled spin pair with $H_S=J_x\sigma_1^x\sigma_2^x+J_y\sigma_1^y\sigma_2^y+h_1\sigma_1^z+h_2\sigma_2^z$ and probe coupling $B=b_1\sigma_1^z+b_2\sigma_2^z$.
Both conserve parity $P=\sigma_1^z\sigma_2^z$, with $w_{x\pm}=J_x\mp J_y$, $w_{z\pm}=h_1\pm h_2$, and $\kappa_\pm=b_1\pm b_2$.
The exchange term $J_x+J_y$ couples the odd-parity states, whereas the anisotropic term $J_x-J_y$ creates or annihilates pairs of excitations in the even sector.
These processes correspond to $i\mathrm{SWAP}$- and $b\mathrm{SWAP}$-type interactions in superconducting qubits~\cite{Roth2017,Wei2024}; independently tunable anisotropic $XY$ couplings have also been demonstrated in trapped ions~\cite{Kotibhaskar2024}.
The odd block also has the exchange-coupled pseudospin structure used for monitored nuclear-spin pairs~\cite{dasari2022purification,pfender2019,jerger2023quench,abobeih2018}.
We use the maximally mixed state $I_2/2$ in each sector, with equal priors $\mu_+=\mu_-=1/2$.

\begin{figure}[tbp]
\includegraphics[width=\linewidth]{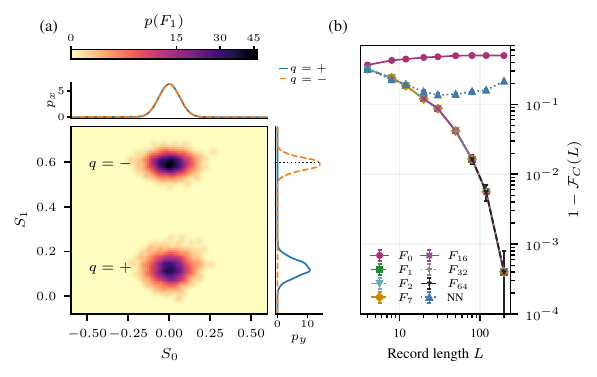}
\caption{Parity readout with equal sector weights and $\phi=\pi/2$, with the odd sector near a Markov point. (a) Joint density $p(F_1)=p(S_0,S_1)$ at $L=1000$, estimated from independent test records, with separately normalized sector-conditioned marginal densities above and to the right. The dotted horizontal line in the right marginal marks the exact Markov prediction $\mathbb E_-[S_1]=2s_--1$ with $s_-=0.8$. The horizontal colorbar indicates the joint density. (b) Classification infidelities of the $F_K$ classifiers with $K=0,1,2,7,16,32,64$ and the full-record NN. All tested $F_K$ curves with $K\ge1$ nearly coincide. Error bars indicate approximate 95\% confidence intervals from finite test sampling.}
\label{fig:Markov}
\end{figure}

For the maximally mixed preparations and $\phi=\pi/2$, both parity sectors have identical single-outcome statistics, with $\mathbb E_q[S_0]=0$, making them indistinguishable from a single probe outcome~\cite{supplementaryMaterial}.
Their temporal correlations, however, can differ because of their distinct internal dynamics.
To illustrate analytically how temporal correlations reveal sector information, we consider a special limit at $w_{zq}=0$ when $(\kappa_q^2/\nu_q^2)\sin^2(\nu_qT)=1/2$, with $\nu_q=\sqrt{w_{xq}^2+\kappa_q^2}$; both Kraus operators in sector $q$ then have rank one~\cite{supplementaryMaterial}.
Each outcome prepares a fixed state, and the record forms a symmetric first-order Markov chain with stay probability $s_q=\Pr_q(x_{n+1}=x_n)=2\cos^2(\nu_qT)$.
In this limit, $\mathbb E_q[S_k]=(2s_q-1)^k$ for $1\le k<L$.
If both sectors satisfy the rank-one condition, their record likelihood ratio depends only on $S_1$, making it sufficient for classification at every finite $L\ge2$~\cite{supplementaryMaterial}.
% Such rank-one condition also provides a reference for measurement tuning, while transition statistics and hidden Markov models can aid parameter estimation and readout calibration~\cite{Martinez2020,Spethmann2025,tsang2012,gammelmark2013,kiilerich2016,pfender2019,Jin2026,chen2026selfconsistent,chen2026disambiguating}.

We next examine classification beyond this analytically solvable limit using full-rank Kraus operators in both parity sectors.
The odd sector is near the rank-one condition to benchmark the Markov prediction, although this proximity is not required for classification~\cite{supplementaryMaterial}.
Their finite-record $S_0$ distributions nearly coincide [Fig.~\ref{fig:Markov}(a)].
In contrast, their adjacent-outcome correlations differ strongly, producing two well-separated peaks along $S_1$.
Consequently, count-only readout performs close to random guessing, whereas the $F_1$ infidelity decreases rapidly with record length [Fig.~\ref{fig:Markov}(b)].
Including longer-lag correlations provides only a small additional improvement over $F_1$ within the tested range.
Despite using complete measurement records, the NN performs substantially worse than the feature-based classifiers with $K\ge1$, particularly at larger $L$.

\textit{Example III: Multi-Sector Readout.---}
We test multi-sector readout in an interacting five-spin dipolar bath, described by
\begin{equation}
H_S=\sum_{j<k}J_{jk}\left[\sigma_j^z\sigma_k^z-(\sigma_j^+\sigma_k^-+\mathrm{H.c.})\right].
\label{eq:manybody}
\end{equation}
The probe coupling $B=\sum_jh_j\sigma_j^z$ and the system Hamiltonian both conserve the total excitation number $Q=\sum_j(1+\sigma_j^z)/2$, defining six sectors $q=0,\ldots,5$, where $\sigma_j^\pm=(\sigma_j^x\pm i\sigma_j^y)/2$. Each sector starts in the maximally mixed state $I_{d_q}/d_q$, where $d_q=\binom{5}{q}$, with equal priors $\mu_q=1/6$. The coupling parameters are given in the Supplemental Material, and the feature-based classifiers and the NN use the same independent training, validation, and test sets~\cite{supplementaryMaterial}.

\begin{figure}[tbp]
\includegraphics[width=\linewidth]{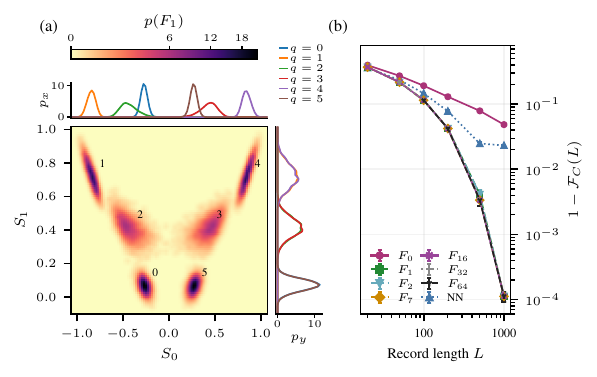}
\caption{Six-sector readout in a five-spin bath with equal sector weights, $T=60\,\mu$s, and $\phi=\pi/2$. (a) Joint density $p(F_1)=p(S_0,S_1)$ at $L=1000$, estimated from 5000 independent test records per sector, with separately normalized sector-conditioned marginal densities above and to the right. Numbers label the sectors. The horizontal colorbar indicates the joint density. (b) Classification infidelities of the $F_K$ classifiers with $K=0,1,2,7,16,32,64$ and the full-record NN. Error bars indicate approximate 95\% confidence intervals from finite test sampling.}
\label{fig:manybody}
\end{figure}

At $L=500$, the classification infidelity decreases by nearly a factor of 20 from $F_0$ to $F_1$ [Fig.~\ref{fig:manybody}(b)], with a smaller additional improvement from $F_7$.
Larger tested values of $K$ provide little further improvement at this record length.
The NN performs slightly better at $L=20$, whereas the feature-based classifiers perform better at the other displayed lengths.
Thus, the benefit of longer-lag correlations depends on the record length and model parameters.
Sectors with overlapping $S_0$ distributions can be distinguished by their different $S_1$ statistics, as seen in the joint distribution [Fig.~\ref{fig:manybody}(a)].
Conversely, charge-conjugate sectors $q$ and $5-q$ have identical pair-correlation means but opposite $S_0$ means, showing that the two statistics provide complementary information.
Spin exchange transports excitations between sites with different probe couplings, producing sector-dependent temporal correlations.

In the Supplemental Material~\cite{supplementaryMaterial}, we also show that classification performance depends nonmonotonically on the RIM evolution time $T$, which affects both single-outcome statistics and temporal correlations.
A six-spin example further demonstrates that longer-lag correlations can substantially improve classification beyond $F_1$.

\textit{Conclusion and outlook.---}
We have introduced a compact, general record feature $F_K$ for conserved-sector identification. It combines the mean outcome with pair correlations over an adjustable lag range, and its evaluation cost grows linearly with record length for fixed $K$. Across the three examples, outcome counts suffice for QND readout, adjacent-pair correlations enable accurate classification near a Markov point, and longer-lag correlations provide additional improvements in interacting spin baths. Increasing $K$ allows us to systematically assess the contribution of temporal correlations at different lags. In the models studied, classification performance rapidly saturates as $K$ increases, showing that accurate classification can be achieved with a compact set of record statistics. This provides a practical guide to choosing the lag range for a given record length and training sample size.

The present framework can be extended in several directions. An important theoretical direction is to reveal the intrinsic relationship between the structure of the monitored system dynamics (e.g., the ranks and commutation relations of the Kraus operators, and the spectrum of the measurement channel) and that of informative record features (e.g., the dominant correlation orders and convergence behaviors). Our work takes only a first step in this direction.
Another direction is to establish classification limits under constraints on measurement time, probe resources, and target disturbance, and to quantify the information lost by compressing complete records into a few features.
Channel spectra and learned dynamical models may guide the joint optimization of measurement protocols and classifiers~\cite{Jin2026,Jirovec2025,McIntyre2026,Cemin2024,gebhart2023learning,ippoliti2024learnability}. Beyond conserved-sector identification, temporal record features may help detect errors and leakage using records from repeated syndrome measurements and estimate unknown parameters in quantum sensing~\cite{bultink2020leakage,Hung2026,Ali2024,tsang2012,kiilerich2016,Ilias2025}. However, experimental studies must also account for readout errors, decoherence, drift, and finite sector lifetimes, requiring appropriate noise mitigation and calibration~\cite{Koh2026,chen2026selfconsistent,chen2026disambiguating,Khanahmadi2023,DAnjou2017,Svastits2026,Xiao2026,Jerger2026,Hazra2025,Dumas2024,Fechant2025}.

\begin{acknowledgments}
This work is supported by the National Natural Science Foundation of China (Grant Nos.~12574082 and E31Q02BG), the Chinese Academy of Sciences (Grant Nos.~E0SEBB11 and E27RBB11), the Quantum Science and Technology---National Science and Technology Major Project (Grant No.~2021ZD0302300), and the Chinese Academy of Sciences Project for Young Scientists in Basic Research (Grant No.~YSBR-090). G.-F.Z. acknowledges support from the National Natural Science Foundation of China (Grant No. 12474353)
\end{acknowledgments}
\bibliography{main}

@article{Jin2026noise,
  title = {Resource-efficient quantum noise spectroscopy with sequential weak measurements},
  author = {Jin, Yuan-De and Ye, Zheng-Fei and Ma, Wen-Long},
  journal = {Phys. Rev. A},
  volume = {114},
  number = {5},
  pages = {052402},
  year = {2026},
  doi = {10.1103/crdc-fzwf},
  url = {https://doi.org/10.1103/crdc-fzwf}
}

@article{Albert2014,
  title = {Symmetries and conserved quantities in {Lindblad} master equations},
  author = {Albert, Victor V. and Jiang, Liang},
  journal = {Phys. Rev. A},
  volume = {89},
  issue = {2},
  pages = {022118},
  numpages = {14},
  year = {2014},
  month = {Feb},
  publisher = {American Physical Society},
  doi = {10.1103/PhysRevA.89.022118},
  url = {https://link.aps.org/doi/10.1103/PhysRevA.89.022118}
}

@article{lidar1998,
  title = {Decoherence-Free Subspaces for Quantum Computation},
  author = {Lidar, D. A. and Chuang, I. L. and Whaley, K. B.},
  year = 1998,
  month = sep,
  journal = {Phys. Rev. Lett.},
  volume = {81},
  number = {12},
  pages = {2594--2597},
  issn = {0031-9007, 1079-7114},
  doi = {10.1103/PhysRevLett.81.2594},
  copyright = {http://link.aps.org/licenses/aps-default-license},
  langid = {english}
}

@article{buca2012,
  title = {A Note on Symmetry Reductions of the {Lindblad} Equation: Transport in Constrained Open Spin Chains},
  shorttitle = {A Note on Symmetry Reductions of the {{Lindblad}} Equation},
  author = {Bu{\v c}a, Berislav and Prosen, Toma{\v z}},
  year = 2012,
  month = jul,
  journal = {New J. Phys.},
  volume = {14},
  number = {7},
  pages = {073007},
  issn = {1367-2630},
  doi = {10.1088/1367-2630/14/7/073007},
}

@article{sanchezMunoz2019,
  title = {Symmetries and Conservation Laws in Quantum Trajectories: Dissipative Freezing},
  shorttitle = {Symmetries and Conservation Laws in Quantum Trajectories},
  author = {S{\'a}nchez Mu{\~n}oz, Carlos and Bu{\v c}a, Berislav and Tindall, Joseph and {Gonz{\'a}lez-Tudela}, Alejandro and Jaksch, Dieter and Porras, Diego},
  year = 2019,
  month = oct,
  journal = {Phys. Rev. A},
  volume = {100},
  number = {4},
  pages = {042113},
  issn = {2469-9926, 2469-9934},
  doi = {10.1103/PhysRevA.100.042113},
  langid = {english},
}

@article{villazon2020,
  title = {Integrability and Dark States in an Anisotropic Central Spin Model},
  author = {Villazon, Tamiro and Chandran, Anushya and Claeys, Pieter W.},
  year = 2020,
  month = aug,
  journal = {Phys. Rev. Res.},
  volume = {2},
  number = {3},
  pages = {032052},
  issn = {2643-1564},
  doi = {10.1103/PhysRevResearch.2.032052},
  langid = {english}
}

@article{bauer2011,
  title = {Convergence of Repeated Quantum Nondemolition Measurements and Wave-Function Collapse},
  author = {Bauer, Michel and Bernard, Denis},
  year = 2011,
  month = oct,
  journal = {Phys. Rev. A},
  volume = {84},
  number = {4},
  pages = {044103},
  issn = {1050-2947, 1094-1622},
  doi = {10.1103/PhysRevA.84.044103},
  copyright = {http://link.aps.org/licenses/aps-default-license},
  langid = {english},
}

@article{bauer2013,
  title = {Repeated Quantum Non-Demolition Measurements: Convergence and Continuous Time Limit},
  shorttitle = {Repeated {{Quantum Non-Demolition Measurements}}},
  author = {Bauer, Michel and Benoist, Tristan and Bernard, Denis},
  year = 2013,
  month = may,
  journal = {Ann. Henri Poincar\'e},
  volume = {14},
  number = {4},
  pages = {639--679},
  issn = {1424-0637, 1424-0661},
  doi = {10.1007/s00023-012-0204-x},
  copyright = {http://www.springer.com/tdm},
  langid = {english}
}

@article{ma2023,
  title = {Sequential Generalized Measurements: Asymptotics, Typicality, and Emergent Projective Measurements},
  shorttitle = {Sequential Generalized Measurements},
  author = {Ma, Wen-Long and Li, Shu-Shen and Liu, Ren-Bao},
  year = 2023,
  month = jan,
  journal = {Phys. Rev. A},
  volume = {107},
  number = {1},
  pages = {012217},
  issn = {2469-9926, 2469-9934},
  doi = {10.1103/PhysRevA.107.012217},
  langid = {english}
}

@article{garrahan2010,
  title = {Thermodynamics of Quantum Jump Trajectories},
  author = {Garrahan, Juan P. and Lesanovsky, Igor},
  year = 2010,
  month = apr,
  journal = {Phys. Rev. Lett.},
  volume = {104},
  number = {16},
  pages = {160601},
  issn = {0031-9007, 1079-7114},
  doi = {10.1103/PhysRevLett.104.160601},
  copyright = {http://link.aps.org/licenses/aps-default-license},
  langid = {english},
}

@article{lesanovsky2013,
  title = {Characterization of Dynamical Phase Transitions in Quantum Jump Trajectories Beyond the Properties of the Stationary State},
  author = {Lesanovsky, Igor and Van Horssen, Merlijn and Gu{\c t}{\u a}, M{\u a}d{\u a}lin and Garrahan, Juan P.},
  year = 2013,
  month = apr,
  journal = {Phys. Rev. Lett.},
  volume = {110},
  number = {15},
  pages = {150401},
  issn = {0031-9007, 1079-7114},
  doi = {10.1103/PhysRevLett.110.150401},
  copyright = {http://link.aps.org/licenses/aps-default-license},
  langid = {english},
}

@article{carollo2019,
  title = {Unraveling the Large Deviation Statistics of {Markovian} Open Quantum Systems},
  author = {Carollo, Federico and Jack, Robert L. and Garrahan, Juan P.},
  year = 2019,
  month = apr,
  journal = {Phys. Rev. Lett.},
  volume = {122},
  number = {13},
  pages = {130605},
  issn = {0031-9007, 1079-7114},
  doi = {10.1103/PhysRevLett.122.130605},
  langid = {english},
}

@article{tsang2012,
  title = {Continuous Quantum Hypothesis Testing},
  author = {Tsang, Mankei},
  year = 2012,
  month = apr,
  journal = {Phys. Rev. Lett.},
  volume = {108},
  number = {17},
  pages = {170502},
  issn = {0031-9007, 1079-7114},
  doi = {10.1103/PhysRevLett.108.170502},
  copyright = {http://link.aps.org/licenses/aps-default-license},
  langid = {english},
}

@article{gammelmark2013,
  title = {{Bayesian} Parameter Inference from Continuously Monitored Quantum Systems},
  author = {Gammelmark, S{\o}ren and M{\o}lmer, Klaus},
  year = 2013,
  month = mar,
  journal = {Phys. Rev. A},
  volume = {87},
  number = {3},
  pages = {032115},
  issn = {1050-2947, 1094-1622},
  doi = {10.1103/PhysRevA.87.032115},
  copyright = {http://link.aps.org/licenses/aps-default-license},
  langid = {english},
}

@article{albert2019,
  title = {Asymptotics of Quantum Channels: Conserved Quantities, an Adiabatic Limit, and Matrix Product States},
  shorttitle = {Asymptotics of Quantum Channels},
  author = {Albert, Victor V.},
  year = 2019,
  month = jun,
  journal = {Quantum},
  volume = {3},
  pages = {151},
  issn = {2521-327X},
  doi = {10.22331/q-2019-06-06-151},
  copyright = {https://creativecommons.org/licenses/by/4.0/},
  langid = {english},
}

@misc{supplementaryMaterial,
 note = {See Supplemental Material at [URL will be inserted by publisher] for conserved-sector derivations, analytical models, and details of classifiers and numerical simulations. It also contains Ref.~\cite{Liu2020}.}
}

@article{ferrari2025dissipative,
  title = {Dissipative Quantum Chaos Unveiled by Stochastic Quantum Trajectories},
  author = {Ferrari, Filippo and Gravina, Luca and Eeltink, Debbie and Scarlino, Pasquale and Savona, Vincenzo and Minganti, Fabrizio},
  year = 2025,
  month = mar,
  journal = {Phys. Rev. Res.},
  volume = {7},
  number = {1},
  pages = {013276},
  issn = {2643-1564},
  doi = {10.1103/PhysRevResearch.7.013276},
  langid = {english},
}

@article{mochizuki2025spectral,
  title = {Measurement-Induced Spectral Transition},
  author = {Mochizuki, Ken and Hamazaki, Ryusuke},
  year = 2025,
  month = jan,
  journal = {Phys. Rev. Lett.},
  volume = {134},
  number = {1},
  pages = {010410},
  issn = {0031-9007, 1079-7114},
  doi = {10.1103/PhysRevLett.134.010410},
  langid = {english},
}

@article{li2025monitored,
  title = {Monitored Long-Range Interacting Systems: Spin-Wave Theory for Quantum Trajectories},
  shorttitle = {Monitored Long-Range Interacting Systems},
  author = {Li, Zejian and Delmonte, Anna and Turkeshi, Xhek and Fazio, Rosario},
  year = 2025,
  month = may,
  journal = {Nat. Commun.},
  volume = {16},
  number = {1},
  pages = {4329},
  issn = {2041-1723},
  doi = {10.1038/s41467-025-59557-w},
  langid = {english}
}

@misc{cea2026large,
  title = {Large Deviations and Conditioned Monitored Quantum Systems: A Tensor Network Approach},
  shorttitle = {Large Deviations and Conditioned Monitored Quantum Systems},
  author = {Cea, Mar{\'i}a and Cech, Marcel and Carollo, Federico and Lesanovsky, Igor and Ba{\~n}uls, Mari Carmen},
  year = 2026,
  publisher = {arXiv},
  doi = {10.48550/ARXIV.2603.24225},
  copyright = {Creative Commons Attribution 4.0 International}
}

@article{brun2002,
  title = {A Simple Model of Quantum Trajectories},
  author = {Brun, Todd A.},
  year = 2002,
  month = jul,
  journal = {Am. J. Phys.},
  volume = {70},
  number = {7},
  pages = {719--737},
  issn = {0002-9505, 1943-2909},
  doi = {10.1119/1.1475328},
  langid = {english}
}

@article{cheung2024faraday,
  title = {Quantum Nonlinear Spectroscopy via Correlations of Weak {Faraday}-Rotation Measurements},
  author = {Cheung, Brian Chung Hang and Liu, Ren-Bao},
  year = 2025,
  month = apr,
  journal = {Adv. Quantum Technol.},
  volume = {8},
  number = {4},
  pages = {2300286},
  issn = {2511-9044, 2511-9044},
  doi = {10.1002/qute.202300286},
  langid = {english}
}

@article{liu2010correlations,
  title = {Dynamics Revealed by Correlations of Time-Distributed Weak Measurements of a Single Spin},
  author = {Liu, R-B and Fung, S-H and Fung, H-K and Korotkov, A N and Sham, L J},
  year = 2010,
  month = jan,
  journal = {New J. Phys.},
  volume = {12},
  number = {1},
  pages = {013018},
  issn = {1367-2630},
}

@article{wiseman1996,
  title = {Quantum Trajectories and Quantum Measurement Theory},
  author = {Wiseman, H M},
  year = 1996,
  month = feb,
  journal = {Quantum Semiclass. Opt.},
  volume = {8},
  number = {1},
  pages = {205--222},
  issn = {1355-5111, 1361-6625},
  doi = {10.1088/1355-5111/8/1/015},
}

@article{sun2014parity,
  title = {Tracking Photon Jumps with Repeated Quantum Non-Demolition Parity Measurements},
  author = {Sun, L. and Petrenko, A. and Leghtas, Z. and Vlastakis, B. and Kirchmair, G. and Sliwa, K. M. and Narla, A. and Hatridge, M. and Shankar, S. and Blumoff, J. and Frunzio, L. and Mirrahimi, M. and Devoret, M. H. and Schoelkopf, R. J.},
  year = 2014,
  month = jul,
  journal = {Nature (London)},
  volume = {511},
  number = {7510},
  pages = {444--448},
  issn = {0028-0836, 1476-4687},
  doi = {10.1038/nature13436},
  langid = {english},
}

@article{pfender2019,
  title = {High-Resolution Spectroscopy of Single Nuclear Spins via Sequential Weak Measurements},
  author = {Pfender, Matthias and Wang, Ping and Sumiya, Hitoshi and Onoda, Shinobu and Yang, Wen and Dasari, Durga Bhaktavatsala Rao and Neumann, Philipp and Pan, Xin-Yu and Isoya, Junichi and Liu, Ren-Bao and Wrachtrup, J{\"o}rg},
  year = 2019,
  month = feb,
  journal = {Nat. Commun.},
  volume = {10},
  number = {1},
  pages = {594},
  issn = {2041-1723},
  doi = {10.1038/s41467-019-08544-z},
  langid = {english}
}

@article{gebhart2023learning,
  title = {Learning Quantum Systems},
  author = {Gebhart, Valentin and Santagati, Raffaele and Gentile, Antonio Andrea and Gauger, Erik M. and Craig, David and Ares, Natalia and Banchi, Leonardo and Marquardt, Florian and Pezz{\`e}, Luca and Bonato, Cristian},
  year = 2023,
  month = feb,
  journal = {Nat. Rev. Phys.},
  issn = {2522-5820},
  doi = {10.1038/s42254-022-00552-1},
  langid = {english}
}

@article{ralph2006qnd,
  title = {Quantum Nondemolition Measurements for Quantum Information},
  author = {Ralph, T. C. and Bartlett, S. D. and O'Brien, J. L. and Pryde, G. J. and Wiseman, H. M.},
  year = 2006,
  month = jan,
  journal = {Phys. Rev. A},
  volume = {73},
  number = {1},
  pages = {012113},
  issn = {1050-2947, 1094-1622},
  doi = {10.1103/PhysRevA.73.012113},
  copyright = {http://link.aps.org/licenses/aps-default-license},
  langid = {english}
}

@article{ma2018phase,
  title = {Phase Transitions in Sequential Weak Measurements},
  author = {Ma, Wen-Long and Wang, Ping and Leong, Weng-Hang and Liu, Ren-Bao},
  year = 2018,
  month = jul,
  journal = {Phys. Rev. A},
  volume = {98},
  number = {1},
  pages = {012117},
  issn = {2469-9926, 2469-9934},
  doi = {10.1103/PhysRevA.98.012117},
  langid = {english},
}

@article{hann2018robust,
  title = {Robust Readout of Bosonic Qubits in the Dispersive Coupling Regime},
  author = {Hann, Connor T. and Elder, Salvatore S. and Wang, Christopher S. and Chou, Kevin and Schoelkopf, Robert J. and Jiang, Liang},
  year = 2018,
  month = aug,
  journal = {Phys. Rev. A},
  volume = {98},
  number = {2},
  pages = {022305},
  issn = {2469-9926, 2469-9934},
  doi = {10.1103/PhysRevA.98.022305},
  langid = {english},
}

@article{wang2023synthesize,
  title = {Using Weak Measurements to Synthesize Projective Measurement of Nonconserved Observables of Weakly Coupled Nuclear Spins},
  author = {Wang, Ping and Yang, Wen and Liu, Renbao},
  year = 2023,
  month = may,
  journal = {Phys. Rev. Applied},
  volume = {19},
  number = {5},
  pages = {054037},
  issn = {2331-7019},
  doi = {10.1103/PhysRevApplied.19.054037},
  langid = {english},
}

@article{jordan2005kicked,
  title = {Quantum Nondemolition Measurement of a Kicked Qubit},
  author = {Jordan, Andrew N. and B{\"u}ttiker, Markus},
  year = 2005,
  month = mar,
  journal = {Phys. Rev. B},
  volume = {71},
  number = {12},
  pages = {125333},
  issn = {1098-0121, 1550-235X},
  doi = {10.1103/PhysRevB.71.125333},
  copyright = {http://link.aps.org/licenses/aps-default-license},
  langid = {english}
}

@article{wang2019arbitrary,
  title = {Characterization of Arbitrary-Order Correlations in Quantum Baths by Weak Measurement},
  author = {Wang, Ping and Chen, Chong and Peng, Xinhua and Wrachtrup, J{\"o}rg and Liu, Ren-Bao},
  year = 2019,
  month = aug,
  journal = {Phys. Rev. Lett.},
  volume = {123},
  number = {5},
  pages = {050603},
  issn = {0031-9007, 1079-7114},
  doi = {10.1103/PhysRevLett.123.050603},
  langid = {english},
}

@article{montenegro2022sequential,
  title = {Sequential Measurements for Quantum-Enhanced Magnetometry in Spin Chain Probes},
  author = {Montenegro, Victor and Jones, Gareth Si{\^o}n and Bose, Sougato and Bayat, Abolfazl},
  year = 2022,
  month = sep,
  journal = {Phys. Rev. Lett.},
  volume = {129},
  number = {12},
  pages = {120503},
  issn = {0031-9007, 1079-7114},
  doi = {10.1103/PhysRevLett.129.120503},
  langid = {english}
}

@article{shen2023quantum,
  title = {Detection of Quantum Signals Free of Classical Noise via Quantum Correlation},
  author = {Shen, Yang and Wang, Ping and Cheung, Chun Tung and Wrachtrup, J{\"o}rg and Liu, Ren-Bao and Yang, Sen},
  year = 2023,
  month = feb,
  journal = {Phys. Rev. Lett.},
  volume = {130},
  number = {7},
  pages = {070802},
  issn = {0031-9007, 1079-7114},
  doi = {10.1103/PhysRevLett.130.070802},
  langid = {english},
}

@article{wu2024selective,
  title = {Selective Detection of Dynamics-Complete Set of Correlations via Quantum Channels},
  author = {Wu, Ze and Wang, Ping and Wang, Tianyun and Li, Yuchen and Liu, Ran and Chen, Yuquan and Peng, Xinhua and Liu, Ren-Bao},
  year = 2024,
  month = may,
  journal = {Phys. Rev. Lett.},
  volume = {132},
  number = {20},
  pages = {200802},
  issn = {0031-9007, 1079-7114},
  doi = {10.1103/PhysRevLett.132.200802},
  langid = {english}
}

@article{marcks2024charge,
  title = {Quantum Spin Probe of Single Charge Dynamics},
  author = {Marcks, Jonathan C. and Onizhuk, Mykyta and Wang, Yu-Xin and Zhu, Yizhi and Jin, Yu and Soloway, Benjamin S. and Fukami, Masaya and Delegan, Nazar and Heremans, F. Joseph and Clerk, Aashish A. and Galli, Giulia and Awschalom, David D.},
  year = 2024,
  month = sep,
  journal = {Phys. Rev. Lett.},
  volume = {133},
  number = {13},
  pages = {130802},
  issn = {0031-9007, 1079-7114},
  doi = {10.1103/PhysRevLett.133.130802},
  langid = {english},
}

@article{elder2020highfidelity,
  title = {High-Fidelity Measurement of Qubits Encoded in Multilevel Superconducting Circuits},
  author = {Elder, Salvatore S. and Wang, Christopher S. and Reinhold, Philip and Hann, Connor T. and Chou, Kevin S. and Lester, Brian J. and Rosenblum, Serge and Frunzio, Luigi and Jiang, Liang and Schoelkopf, Robert J.},
  year = 2020,
  month = jan,
  journal = {Phys. Rev. X},
  volume = {10},
  number = {1},
  pages = {011001},
  issn = {2160-3308},
  doi = {10.1103/PhysRevX.10.011001},
  urldate = {2026-11-06},
  langid = {english}
}

@article{jerger2023quench,
  title = {Detecting Spin-Bath Polarization with Quantum Quench Phase Shifts of Single Spins in Diamond},
  author = {Jerger, Paul C. and Wang, Yu-Xin and Onizhuk, Mykyta and Soloway, Benjamin S. and Solomon, Michael T. and Egerstrom, Christopher and Heremans, F. Joseph and Galli, Giulia and Clerk, Aashish A. and Awschalom, David D.},
  year = 2023,
  month = oct,
  journal = {PRX Quantum},
  volume = {4},
  number = {4},
  pages = {040315},
  issn = {2691-3399},
  doi = {10.1103/PRXQuantum.4.040315},
  langid = {english}
}

@article{ivashkov2024dynamic,
  title = {High-Fidelity, Multiqubit Generalized Measurements with Dynamic Circuits},
  author = {Ivashkov, Petr and Uchehara, Gideon and Jiang, Liang and Wang, Derek S. and Seif, Alireza},
  year = 2024,
  month = jul,
  journal = {PRX Quantum},
  volume = {5},
  number = {3},
  pages = {030315},
  issn = {2691-3399},
  doi = {10.1103/PRXQuantum.5.030315},
  langid = {english}
}

@article{chen2026selfconsistent,
  title = {Efficient Self-Consistent Learning of Gate Set {Pauli} Noise},
  author = {Chen, Senrui and Zhang, Zhihan and Jiang, Liang and Flammia, Steven T.},
  year = 2026,
  month = jan,
  journal = {PRX Quantum},
  volume = {7},
  number = {1},
  pages = {010305},
  issn = {2691-3399},
  doi = {10.1103/1pnv-t9px},
  langid = {english}
}

@article{chen2026disambiguating,
  title = {Disambiguating {Pauli} Noise in Quantum Computers},
  author = {Chen, Edward H. and Chen, Senrui and Fischer, Laurin E. and Eddins, Andrew and Govia, Luke C. G. and Mitchell, Brad and He, Andre and Kim, Youngseok and Jiang, Liang and Seif, Alireza},
  year = 2026,
  month = sep,
  journal = {PRX Quantum},
  volume = {7},
  number = {3},
  pages = {033045},
  issn = {2691-3399},
  doi = {10.1103/69wc-gzl6},
  langid = {english},
}

@article{zhan2024conservation,
  title = {Learning Conservation Laws in Unknown Quantum Dynamics},
  author = {Zhan, Yongtao and Elben, Andreas and Huang, Hsin-Yuan and Tong, Yu},
  year = 2024,
  month = mar,
  journal = {PRX Quantum},
  volume = {5},
  number = {1},
  pages = {010350},
  issn = {2691-3399},
  doi = {10.1103/PRXQuantum.5.010350},
  langid = {english}
}

@article{ippoliti2024learnability,
  title = {Learnability Transitions in Monitored Quantum Dynamics via Eavesdropper's Classical Shadows},
  author = {Ippoliti, Matteo and Khemani, Vedika},
  year = 2024,
  month = apr,
  journal = {PRX Quantum},
  volume = {5},
  number = {2},
  pages = {020304},
  issn = {2691-3399},
  doi = {10.1103/PRXQuantum.5.020304},
  langid = {english},
}

@article{barratt2022learnability,
  title = {Transitions in the Learnability of Global Charges from Local Measurements},
  author = {Barratt, Fergus and Agrawal, Utkarsh and Potter, Andrew C. and Gopalakrishnan, Sarang and Vasseur, Romain},
  year = 2022,
  month = nov,
  journal = {Phys. Rev. Lett.},
  volume = {129},
  number = {20},
  pages = {200602},
  issn = {0031-9007, 1079-7114},
  doi = {10.1103/PhysRevLett.129.200602},
  langid = {english},
}

@article{abobeih2018,
  title = {One-Second Coherence for a Single Electron Spin Coupled to a Multi-Qubit Nuclear-Spin Environment},
  author = {Abobeih, M. H. and Cramer, J. and Bakker, M. A. and Kalb, N. and Markham, M. and Twitchen, D. J. and Taminiau, T. H.},
  year = 2018,
  month = jun,
  journal = {Nat. Commun.},
  volume = {9},
  number = {1},
  pages = {2552},
  issn = {2041-1723},
  doi = {10.1038/s41467-018-04916-z},
  langid = {english}
}

@misc{dong2026,
  title = {Quantum Probe Advantage in Learning Many-Body Systems},
  author = {Dong, Wenzheng and Green, Andrew G. and Vedral, Vlatko and Sun, Jinzhao},
  year = 2026,
  month = jul,
  publisher = {arXiv},
  doi = {10.48550/ARXIV.2607.11829},
  copyright = {Creative Commons Attribution 4.0 International}
}

@misc{Jirovec2025,
 title = {Many-body interferometry with semiconductor spins},
 author = {Jirovec, Daniel and Reale, Stefano and Cova-Fari{\~{n}}a, Pablo and Ventura-Meinersen, Christian and Nguyen, Minh T. P. and Zhang, Xin and Oosterhout, Stefan D. and Scappucci, Giordano and Veldhorst, Menno and Rimbach-Russ, Maximilian and Bosco, Stefano and Vandersypen, Lieven M. K.},
 year = {2025},
 eprint = {2511.04310},
 archivePrefix = {arXiv},
 url = {https://arxiv.org/abs/2511.04310},
}

@misc{Hung2026,
 title = {Fast, High-Fidelity Erasure Detection of Dual-Rail Qubits with Symmetrically Coupled Readout},
 author = {Hung, Jimmy Shih-Chun and Haim, Arbel and Raha, Mouktik and Kim, Gihwan and Huang, Ziwen and Chou, Ming-Han and D'Ewart, Mitch and Davis, Erik and Mishra, Anurag and Arriola, Patricio Arrangoiz and Khalajhedayati, Amirhossein and Hover, David and Brand{\~{a}}o, Fernando G. S. L. and Clerk, Aashish A. and Retzker, Alex and Levine, Harry and Painter, Oskar},
 year = {2026},
 eprint = {2604.16292},
 archivePrefix = {arXiv},
 url = {https://arxiv.org/abs/2604.16292},
}

@misc{McIntyre2026,
 title = {One-clean-qubit spectroscopy of simulated {Kitaev} chains},
 author = {McIntyre, Z. M. and Loss, Daniel},
 year = {2026},
 eprint = {2609.11513},
 archivePrefix = {arXiv},
 url = {https://arxiv.org/abs/2609.11513},
}

@incollection{Girvin2014,
  title = {Circuit {QED}: superconducting qubits coupled to microwave photons},
  author = {Girvin, S. M.},
  year = {2014},
  pages = {113-256},
  doi = {10.1093/acprof:oso/9780199681181.003.0003},
  url = {https://doi.org/10.1093/acprof:oso/9780199681181.003.0003},
  booktitle = {Quantum Machines: Measurement and Control of Engineered Quantum Systems},
  publisher = {Oxford University Press},
}

@article{Liu2020,
  title = {Repetitive readout enhanced by machine learning},
  author = {Liu, Genyue and Chen, Mo and Liu, Yi-Xiang and Layden, David and Cappellaro, Paola},
  journal = {Mach. Learn.: Sci. Technol.},
  year = {2020},
  volume = {1},
  number = {1},
  pages = {015003},
  doi = {10.1088/2632-2153/ab4e24},
  url = {https://doi.org/10.1088/2632-2153/ab4e24},
}

@article{Majaniemi2026,
  title = {Reducing quantum error correction overhead using soft information},
  author = {Majaniemi, Joonas and Matekole, Elisha S},
  journal = {Quantum Sci. Technol.},
  year = {2026},
  volume = {11},
  number = {2},
  pages = {025024},
  doi = {10.1088/2058-9565/ae4e55},
  url = {https://doi.org/10.1088/2058-9565/ae4e55},
}

@article{Martinez2020,
  title = {Improving qubit readout with hidden {Markov} models},
  author = {Martinez, Luis A. and Rosen, Yaniv J. and DuBois, Jonathan L.},
  journal = {Phys. Rev. A},
  year = {2020},
  volume = {102},
  number = {6},
  pages = {062426},
  doi = {10.1103/physreva.102.062426},
  url = {https://doi.org/10.1103/physreva.102.062426},
}

@article{Debnath2021,
  title = {Ancilla-mediated qubit readout and heralded entanglement between rare-earth dopant ions in crystals},
  author = {Debnath, Kamanasish and Kiilerich, Alexander Holm and M{\o}lmer, Klaus},
  journal = {Phys. Rev. A},
  year = {2021},
  volume = {103},
  number = {4},
  pages = {043705},
  doi = {10.1103/physreva.103.043705},
  url = {https://doi.org/10.1103/physreva.103.043705},
}

@article{Khanahmadi2023,
  title = {Qubit readout and quantum sensing with pulses of quantum radiation},
  author = {Khanahmadi, Maryam and M{\o}lmer, Klaus},
  journal = {Phys. Rev. A},
  year = {2023},
  volume = {107},
  number = {1},
  pages = {013705},
  doi = {10.1103/physreva.107.013705},
  url = {https://doi.org/10.1103/physreva.107.013705},
}

@article{Cemin2024,
  title = {Machine learning of reduced quantum channels on noisy intermediate-scale quantum devices},
  author = {Cemin, Giovanni and Cech, Marcel and Weiss, Erik and Soltan, Stanislaw and Braun, Daniel and Lesanovsky, Igor and Carollo, Federico},
  journal = {Phys. Rev. A},
  year = {2024},
  volume = {110},
  number = {5},
  pages = {052418},
  doi = {10.1103/physreva.110.052418},
  url = {https://doi.org/10.1103/physreva.110.052418},
}

@article{Ilias2025,
  title = {Biasing quantum trajectories for enhanced sensing},
  author = {Ilias, Theodoros},
  journal = {Phys. Rev. A},
  year = {2025},
  volume = {111},
  number = {4},
  pages = {042432},
  doi = {10.1103/physreva.111.042432},
  url = {https://doi.org/10.1103/physreva.111.042432},
}

@article{Jin2026,
  title = {Spectrum measurement of quantum channels and application to {Hamiltonian} parameter estimation},
  author = {Jin, Yuan-De and Ma, Wen-Long},
  journal = {Phys. Rev. A},
  year = {2026},
  volume = {113},
  number = {1},
  pages = {L010406},
  doi = {10.1103/66yg-qs8x},
  url = {https://doi.org/10.1103/66yg-qs8x},
}

@article{Cech2026,
  title = {Revealing emergent many-body phenomena by analyzing large-scale space-time records of monitored quantum systems},
  author = {Cech, Marcel and De Fazio, Cecilia and Cea, Mar{\'{i}}a and Ba{\~{n}}uls, Mari Carmen and Lesanovsky, Igor and Carollo, Federico},
  journal = {Phys. Rev. A},
  year = {2026},
  volume = {113},
  number = {6},
  pages = {L060201},
  doi = {10.1103/flpc-1sls},
  url = {https://doi.org/10.1103/flpc-1sls},
}

@article{DAnjou2014a,
  title = {Optimal post-processing for a generic single-shot qubit readout},
  author = {D'Anjou, B. and Coish, W. A.},
  journal = {Phys. Rev. A},
  year = {2014},
  volume = {89},
  number = {1},
  pages = {012313},
  doi = {10.1103/physreva.89.012313},
  url = {https://doi.org/10.1103/physreva.89.012313},
}

@article{DAnjou2017,
  title = {Enhancing qubit readout through dissipative sub-{Poissonian} dynamics},
  author = {D'Anjou, B. and Coish, W. A.},
  journal = {Phys. Rev. A},
  year = {2017},
  volume = {96},
  number = {5},
  pages = {052321},
  doi = {10.1103/physreva.96.052321},
  url = {https://doi.org/10.1103/physreva.96.052321},
}

@article{Ali2024,
  title = {Reducing the error rate of a superconducting logical qubit using analog readout information},
  author = {Ali, Hany and Marques, Jorge and Crawford, Ophelia and Majaniemi, Joonas and Serra-Peralta, Marc and Byfield, David and Varbanov, Boris and Terhal, Barbara M. and DiCarlo, Leonardo and Campbell, Earl T.},
  journal = {Phys. Rev. Applied},
  year = {2024},
  volume = {22},
  number = {4},
  pages = {044031},
  doi = {10.1103/physrevapplied.22.044031},
  url = {https://doi.org/10.1103/physrevapplied.22.044031},
}

@article{DAnjou2019,
  title = {Optimal dispersive readout of a spin qubit with a microwave resonator},
  author = {D'Anjou, B. and Burkard, Guido},
  journal = {Phys. Rev. B},
  year = {2019},
  volume = {100},
  number = {24},
  pages = {245427},
  doi = {10.1103/physrevb.100.245427},
  url = {https://doi.org/10.1103/physrevb.100.245427},
}

@article{Ginzel2023,
  title = {Simultaneous transient dispersive readout of multiple spin qubits},
  author = {Ginzel, Florian and Burkard, Guido},
  journal = {Phys. Rev. B},
  year = {2023},
  volume = {108},
  number = {12},
  pages = {125437},
  doi = {10.1103/physrevb.108.125437},
  url = {https://doi.org/10.1103/physrevb.108.125437},
}

@article{Spethmann2025,
  title = {Spin-qubit readout analysis based on a hidden {Markov} model},
  author = {Spethmann, Maria and Stano, Peter and Loss, Daniel},
  journal = {Phys. Rev. B},
  year = {2025},
  volume = {112},
  number = {11},
  pages = {115304},
  doi = {10.1103/x8yx-5111},
  url = {https://doi.org/10.1103/x8yx-5111},
}

@article{Li2026,
  title = {Exact large deviations and emergent long-range correlations in sequential quantum {East} circuits},
  author = {Li, Jimin and Bertini, Bruno and Garrahan, Juan P. and Jack, Robert L.},
  journal = {Phys. Rev. B},
  year = {2026},
  volume = {113},
  number = {10},
  pages = {L100305},
  doi = {10.1103/72kx-56lt},
  url = {https://doi.org/10.1103/72kx-56lt},
}

@article{Vijay2011,
  title = {Observation of Quantum Jumps in a Superconducting Artificial Atom},
  author = {Vijay, R. and Slichter, D. H. and Siddiqi, I.},
  journal = {Phys. Rev. Lett.},
  year = {2011},
  volume = {106},
  number = {11},
  pages = {110502},
  doi = {10.1103/physrevlett.106.110502},
  url = {https://doi.org/10.1103/physrevlett.106.110502},
}

@article{Lee2012,
  title = {Collective Quantum Jumps of {Rydberg} Atoms},
  author = {Lee, Tony E. and H{\"{a}}ffner, H. and Cross, M. C.},
  journal = {Phys. Rev. Lett.},
  year = {2012},
  volume = {108},
  number = {2},
  pages = {023602},
  doi = {10.1103/physrevlett.108.023602},
  url = {https://doi.org/10.1103/physrevlett.108.023602},
}

@article{Flindt2013,
  title = {Trajectory Phase Transitions, {Lee}-{Yang} Zeros, and High-Order Cumulants in Full Counting Statistics},
  author = {Flindt, Christian and Garrahan, Juan P.},
  journal = {Phys. Rev. Lett.},
  year = {2013},
  volume = {110},
  number = {5},
  pages = {050601},
  doi = {10.1103/physrevlett.110.050601},
  url = {https://doi.org/10.1103/physrevlett.110.050601},
}

@article{Znidaric2014,
  title = {Exact Large-Deviation Statistics for a Nonequilibrium Quantum Spin Chain},
  author = {{\v{Z}}nidari{\v{c}}, Marko},
  journal = {Phys. Rev. Lett.},
  year = {2014},
  volume = {112},
  number = {4},
  pages = {040602},
  doi = {10.1103/physrevlett.112.040602},
  url = {https://doi.org/10.1103/physrevlett.112.040602},
}

@article{DAnjou2014b,
  title = {Soft Decoding of a Qubit Readout Apparatus},
  author = {D'Anjou, B. and Coish, W. A.},
  journal = {Phys. Rev. Lett.},
  year = {2014},
  volume = {113},
  number = {23},
  pages = {230402},
  doi = {10.1103/physrevlett.113.230402},
  url = {https://doi.org/10.1103/physrevlett.113.230402},
}

@article{Ashida2018,
  title = {Thermalization and Heating Dynamics in Open Generic Many-Body Systems},
  author = {Ashida, Yuto and Saito, Keiji and Ueda, Masahito},
  journal = {Phys. Rev. Lett.},
  year = {2018},
  volume = {121},
  number = {17},
  pages = {170402},
  doi = {10.1103/physrevlett.121.170402},
  url = {https://doi.org/10.1103/physrevlett.121.170402},
}

@article{Chen2021,
  title = {Quantum Jumps in the Non-{Hermitian} Dynamics of a Superconducting Qubit},
  author = {Chen, Weijian and Abbasi, Maryam and Joglekar, Yogesh N. and Murch, Kater W.},
  journal = {Phys. Rev. Lett.},
  year = {2021},
  volume = {127},
  number = {14},
  pages = {140504},
  doi = {10.1103/physrevlett.127.140504},
  url = {https://doi.org/10.1103/physrevlett.127.140504},
}

@article{Kleinherbers2022,
  title = {Pushing the Limits in Real-Time Measurements of Quantum Dynamics},
  author = {Kleinherbers, E. and Stegmann, P. and Kurzmann, A. and Geller, M. and Lorke, A. and K{\"{o}}nig, J.},
  journal = {Phys. Rev. Lett.},
  year = {2022},
  volume = {128},
  number = {8},
  pages = {087701},
  doi = {10.1103/physrevlett.128.087701},
  url = {https://doi.org/10.1103/physrevlett.128.087701},
}

@article{Cech2023,
  title = {Thermodynamics of Quantum Trajectories on a Quantum Computer},
  author = {Cech, Marcel and Lesanovsky, Igor and Carollo, Federico},
  journal = {Phys. Rev. Lett.},
  year = {2023},
  volume = {131},
  number = {12},
  pages = {120401},
  doi = {10.1103/physrevlett.131.120401},
  url = {https://doi.org/10.1103/physrevlett.131.120401},
}

@article{Paulino2024,
  title = {Large Deviation Full Counting Statistics in Adiabatic Open Quantum Dynamics},
  author = {Paulino, Paulo J. and Lesanovsky, Igor and Carollo, Federico},
  journal = {Phys. Rev. Lett.},
  year = {2024},
  volume = {132},
  number = {26},
  pages = {260402},
  doi = {10.1103/physrevlett.132.260402},
  url = {https://doi.org/10.1103/physrevlett.132.260402},
}

@article{Hazra2025,
  title = {Benchmarking the Readout of a Superconducting Qubit for Repeated Measurements},
  author = {Hazra, S. and Dai, W. and Connolly, T. and Kurilovich, P. D. and Wang, Z. and Frunzio, L. and Devoret, M. H.},
  journal = {Phys. Rev. Lett.},
  year = {2025},
  volume = {134},
  number = {10},
  pages = {100601},
  doi = {10.1103/physrevlett.134.100601},
  url = {https://doi.org/10.1103/physrevlett.134.100601},
}

@article{Jin2025,
  title = {General Approach to Error Detection of Bosonic Codes via Phase Estimation},
  author = {Jin, Yuan-De and Zhang, Shi-Yu and Andersen, Ulrik L. and Ma, Wen-Long},
  journal = {Phys. Rev. Lett.},
  year = {2025},
  volume = {135},
  number = {14},
  pages = {140603},
  doi = {10.1103/h6rv-hz81},
  url = {https://doi.org/10.1103/h6rv-hz81},
}

@article{Fechant2025,
  title = {Offset Charge Dependence of Measurement-Induced Transitions in Transmons},
  author = {F{\'{e}}chant, Mathieu and Dumas, Marie Fr{\'{e}}d{\'{e}}rique and B{\'{e}}n{\^{a}}tre, Denis and Gosling, Nicolas and Lenhard, Philipp and Spiecker, Martin and Geisert, Simon and Ihssen, S{\"{o}}ren and Wernsdorfer, Wolfgang and D'Anjou, Benjamin and Blais, Alexandre and Pop, Ioan M.},
  journal = {Phys. Rev. Lett.},
  year = {2025},
  volume = {135},
  number = {18},
  pages = {180603},
  doi = {10.1103/yljv-b4kj},
  url = {https://doi.org/10.1103/yljv-b4kj},
}

@article{Xiao2026,
  title = {Flexible Readout and Unconditional Reset for Superconducting Multiqubit Processors with Tunable {Purcell} Filters},
  author = {Xiao, Yong-Xi and Feng, Da'er and Gu, Xu-Yang and Liang, Gui-Han and Wang, Ming-Chuan and Peng, Zhen-Yu and Chen, Bing-Jie and Yan, Yu and Mei, Zheng-Yang and Zhao, Si-Lu and Bu, Yi-Zhou and Deng, Cheng-Lin and Yang, Kai and Tian, Ye and Song, Xiaohui and Zheng, Dongning and Zhang, Yu-Xiang and Shi, Yun-Hao and Xiang, Zhongcheng and Xu, Kai and Fan, Heng},
  journal = {Phys. Rev. Lett.},
  year = {2026},
  volume = {136},
  number = {7},
  pages = {070601},
  doi = {10.1103/vwrv-x1kr},
  url = {https://doi.org/10.1103/vwrv-x1kr},
}

@article{Svastits2026,
  title = {Readout Sweet Spots for Spin Qubits with Strong Spin-Orbit Interaction},
  author = {Svastits, Domonkos and Het{\'{e}}nyi, Bence and Sz{\'{e}}chenyi, G{\'{a}}bor and Wootton, James and Loss, Daniel and Bosco, Stefano and P{\'{a}}lyi, Andr{\'{a}}s},
  journal = {Phys. Rev. Lett.},
  year = {2026},
  volume = {136},
  number = {11},
  pages = {117001},
  doi = {10.1103/4x97-np1f},
  url = {https://doi.org/10.1103/4x97-np1f},
}

@article{Lee2026,
  title = {Rapid Cavity-Based Midcircuit Measurement and Feedforward in a Neutral Atom Array},
  author = {Lee, Tsai-Chen and Ho, Jacquelyn and Lu, Yue-Hui and Xiang, Tai and Vilas, Nathaniel B. and Yan, Zhenjie and Stamper-Kurn, Dan M.},
  journal = {Phys. Rev. Lett.},
  year = {2026},
  volume = {137},
  number = {20},
  pages = {203601},
  doi = {10.1103/q8k6-mj9z},
  url = {https://doi.org/10.1103/q8k6-mj9z},
}

@article{Cirac1993,
  title = {Preparation of {Fock} states by observation of quantum jumps in an ion trap},
  author = {Cirac, J. I. and Blatt, R. and Parkins, A. S. and Zoller, P.},
  journal = {Phys. Rev. Lett.},
  year = {1993},
  volume = {70},
  number = {6},
  pages = {762-765},
  doi = {10.1103/physrevlett.70.762},
  url = {https://doi.org/10.1103/physrevlett.70.762},
}

@article{Brun1997,
  title = {Quantum Jumps as Decoherent Histories},
  author = {Brun, Todd A.},
  journal = {Phys. Rev. Lett.},
  year = {1997},
  volume = {78},
  number = {10},
  pages = {1833-1837},
  doi = {10.1103/physrevlett.78.1833},
  url = {https://doi.org/10.1103/physrevlett.78.1833},
}

@article{Friesen2004,
  title = {Spin Readout and Initialization in a Semiconductor Quantum Dot},
  author = {Friesen, Mark and Tahan, Charles and Joynt, Robert and Eriksson, M. A.},
  journal = {Phys. Rev. Lett.},
  year = {2004},
  volume = {92},
  number = {3},
  pages = {037901},
  doi = {10.1103/physrevlett.92.037901},
  url = {https://doi.org/10.1103/physrevlett.92.037901},
}

@article{Hilder2022,
  title = {Fault-Tolerant Parity Readout on a Shuttling-Based Trapped-Ion Quantum Computer},
  author = {Hilder, J. and Pijn, D. and Onishchenko, O. and Stahl, A. and Orth, M. and Lekitsch, B. and Rodriguez-Blanco, A. and M{\"{u}}ller, M. and Schmidt-Kaler, F. and Poschinger, U. G.},
  journal = {Phys. Rev. X},
  year = {2022},
  volume = {12},
  number = {1},
  pages = {011032},
  doi = {10.1103/physrevx.12.011032},
  url = {https://doi.org/10.1103/physrevx.12.011032},
}

@article{Oakes2023,
  title = {Fast High-Fidelity Single-Shot Readout of Spins in Silicon Using a Single-Electron Box},
  author = {Oakes, G. A. and Ciriano-Tejel, V. N. and Wise, D. F. and Fogarty, M. A. and Lundberg, T. and Lain{\'{e}}, C. and Schaal, S. and Martins, F. and Ibberson, D. J. and Hutin, L. and Bertrand, B. and Stelmashenko, N. and Robinson, J. W. A. and Ibberson, L. and Hashim, A. and Siddiqi, I. and Lee, A. and Vinet, M. and Smith, C. G. and Morton, J. J. L. and Gonzalez-Zalba, M. F.},
  journal = {Phys. Rev. X},
  year = {2023},
  volume = {13},
  number = {1},
  pages = {011023},
  doi = {10.1103/physrevx.13.011023},
  url = {https://doi.org/10.1103/physrevx.13.011023},
}

@article{Rosenthal2024,
  title = {Single-Shot Readout and Weak Measurement of a Tin-Vacancy Qubit in Diamond},
  author = {Rosenthal, Eric I. and Biswas, Souvik and Scuri, Giovanni and Lee, Hope and Stein, Abigail J. and Kleidermacher, Hannah C. and Grzesik, Jakob and Rugar, Alison E. and Aghaeimeibodi, Shahriar and Riedel, Daniel and Titze, Michael and Bielejec, Edward S. and Choi, Joonhee and Anderson, Christopher P. and Vu{\v{c}}kovi{\'{c}}, Jelena},
  journal = {Phys. Rev. X},
  year = {2024},
  volume = {14},
  number = {4},
  pages = {041008},
  doi = {10.1103/physrevx.14.041008},
  url = {https://doi.org/10.1103/physrevx.14.041008},
}

@article{Dumas2024,
  title = {Measurement-Induced Transmon Ionization},
  author = {Dumas, Marie Fr{\'{e}}d{\'{e}}rique and Groleau-Par{\'{e}}, Benjamin and McDonald, Alexander and Mu{\~{n}}oz-Arias, Manuel H. and Lled{\'{o}}, Crist{\'{o}}bal and D'Anjou, Benjamin and Blais, Alexandre},
  journal = {Phys. Rev. X},
  year = {2024},
  volume = {14},
  number = {4},
  pages = {041023},
  doi = {10.1103/physrevx.14.041023},
  url = {https://doi.org/10.1103/physrevx.14.041023},
}

@article{DAnjou2016,
  title = {Maximal Adaptive-Decision Speedups in Quantum-State Readout},
  author = {D'Anjou, B. and Kuret, L. and Childress, L. and Coish, W. A.},
  journal = {Phys. Rev. X},
  year = {2016},
  volume = {6},
  number = {1},
  pages = {011017},
  doi = {10.1103/physrevx.6.011017},
  url = {https://doi.org/10.1103/physrevx.6.011017},
}

@article{HarveyCollard2018,
  title = {High-Fidelity Single-Shot Readout for a Spin Qubit via an Enhanced Latching Mechanism},
  author = {Harvey-Collard, Patrick and D'Anjou, Benjamin and Rudolph, Martin and Jacobson, N. Tobias and Dominguez, Jason and Ten Eyck, Gregory A. and Wendt, Joel R. and Pluym, Tammy and Lilly, Michael P. and Coish, William A. and Pioro-Ladri{\`{e}}re, Michel and Carroll, Malcolm S.},
  journal = {Phys. Rev. X},
  year = {2018},
  volume = {8},
  number = {2},
  pages = {021046},
  doi = {10.1103/physrevx.8.021046},
  url = {https://doi.org/10.1103/physrevx.8.021046},
}

@article{CirianoTejel2021,
  title = {Spin Readout of a {CMOS} Quantum Dot by Gate Reflectometry and Spin-Dependent Tunneling},
  author = {Ciriano-Tejel, Virginia N. and Fogarty, Michael A. and Schaal, Simon and Hutin, Louis and Bertrand, Benoit and Ibberson, Lisa and Gonzalez-Zalba, M. Fernando and Li, Jing and Niquet, Yann-Michel and Vinet, Maud and Morton, John J.L.},
  journal = {PRX Quantum},
  year = {2021},
  volume = {2},
  number = {1},
  pages = {010353},
  doi = {10.1103/prxquantum.2.010353},
  url = {https://doi.org/10.1103/prxquantum.2.010353},
}

@article{Mielke2021,
  title = {Nuclear Spin Readout in a Cavity-Coupled Hybrid Quantum Dot-Donor System},
  author = {Mielke, Jonas and Petta, Jason R. and Burkard, Guido},
  journal = {PRX Quantum},
  year = {2021},
  volume = {2},
  number = {2},
  pages = {020347},
  doi = {10.1103/prxquantum.2.020347},
  url = {https://doi.org/10.1103/prxquantum.2.020347},
}

@article{Gachter2022,
  title = {Single-Shot Spin Readout in Graphene Quantum Dots},
  author = {G{\"{a}}chter, Lisa Maria and Garreis, Rebekka and Gerber, Jonas Daniel and Ruckriegel, Max Josef and Tong, Chuyao and Kratochwil, Benedikt and de Vries, Folkert Kornelis and Kurzmann, Annika and Watanabe, Kenji and Taniguchi, Takashi and Ihn, Thomas and Ensslin, Klaus and Huang, Wister Wei},
  journal = {PRX Quantum},
  year = {2022},
  volume = {3},
  number = {2},
  pages = {020343},
  doi = {10.1103/prxquantum.3.020343},
  url = {https://doi.org/10.1103/prxquantum.3.020343},
}

@article{Nurizzo2023,
  title = {Complete Readout of Two-Electron Spin States in a Double Quantum Dot},
  author = {Nurizzo, Martin and Jadot, Baptiste and Mortemousque, Pierre-Andr{\'{e}} and Thiney, Vivien and Chanrion, Emmanuel and Niegemann, David and Dartiailh, Matthieu and Ludwig, Arne and Wieck, Andreas D. and B{\"{a}}uerle, Christopher and Urdampilleta, Matias and Meunier, Tristan},
  journal = {PRX Quantum},
  year = {2023},
  volume = {4},
  number = {1},
  pages = {010329},
  doi = {10.1103/prxquantum.4.010329},
  url = {https://doi.org/10.1103/prxquantum.4.010329},
}

@article{Huie2023,
  title = {Repetitive Readout and Real-Time Control of Nuclear Spin Qubits in 
{171Yb}
 Atoms},
  author = {Huie, William and Li, Lintao and Chen, Neville and Hu, Xiye and Jia, Zhubing and Sun, Won Kyu Calvin and Covey, Jacob P.},
  journal = {PRX Quantum},
  year = {2023},
  volume = {4},
  number = {3},
  pages = {030337},
  doi = {10.1103/prxquantum.4.030337},
  url = {https://doi.org/10.1103/prxquantum.4.030337},
}

@article{Serrano2024,
  title = {Improved Single-Shot Qubit Readout Using Twin rf-{SET} Charge Correlations},
  author = {Serrano, Santiago and Feng, MengKe and Lim, Wee Han and Seedhouse, Amanda E. and Tanttu, Tuomo and Gilbert, Will and Escott, Christopher C. and Abrosimov, Nikolay V. and Pohl, Hans-Joachim and Thewalt, Michael L.W. and Hudson, Fay E. and Saraiva, Andre and Dzurak, Andrew S. and Laucht, Arne},
  journal = {PRX Quantum},
  year = {2024},
  volume = {5},
  number = {1},
  pages = {010301},
  doi = {10.1103/prxquantum.5.010301},
  url = {https://doi.org/10.1103/prxquantum.5.010301},
}

@article{Berent2024,
  title = {Analog Information Decoding of Bosonic Quantum Low-Density Parity-Check Codes},
  author = {Berent, Lucas and Hillmann, Timo and Eisert, Jens and Wille, Robert and Roffe, Joschka},
  journal = {PRX Quantum},
  year = {2024},
  volume = {5},
  number = {2},
  pages = {020349},
  doi = {10.1103/prxquantum.5.020349},
  url = {https://doi.org/10.1103/prxquantum.5.020349},
}

@article{Koh2026,
  title = {Readout Error Mitigation for Mid-Circuit Measurements and Feedforward},
  author = {Koh, Jin Ming and Koh, Dax Enshan and Thompson, Jayne},
  journal = {PRX Quantum},
  year = {2026},
  volume = {7},
  number = {1},
  pages = {010317},
  doi = {10.1103/cj89-4h5t},
  url = {https://doi.org/10.1103/cj89-4h5t},
}

@article{Jerger2026,
  title = {Dispersive Qubit Readout with Intrinsic Resonator Reset},
  author = {Jerger, M. and Motzoi, F. and Gao, Y. and Dickel, C. and Buchmann, L. and C{\'{a}}rdenas-L{\'{o}}pez, F. A. and Bengtsson, A. and Tancredi, G. and Warren, C. W. and Bylander, J. and DiVincenzo, D. and Barends, R. and Bushev, P. A.},
  journal = {PRX Quantum},
  year = {2026},
  volume = {7},
  number = {3},
  pages = {033042},
  doi = {10.1103/z5x1-2mr9},
  url = {https://doi.org/10.1103/z5x1-2mr9},
}

@article{Blais2021,
  title = {Circuit quantum electrodynamics},
  author = {Blais, Alexandre and Grimsmo, Arne L. and Girvin, S. M. and Wallraff, Andreas},
  journal = {Rev. Mod. Phys.},
  year = {2021},
  volume = {93},
  number = {2},
  pages = {025005},
  doi = {10.1103/revmodphys.93.025005},
  url = {https://doi.org/10.1103/revmodphys.93.025005},
}

@article{Burkard2023,
  title = {Semiconductor spin qubits},
  author = {Burkard, Guido and Ladd, Thaddeus D. and Pan, Andrew and Nichol, John M. and Petta, Jason R.},
  journal = {Rev. Mod. Phys.},
  year = {2023},
  volume = {95},
  number = {2},
  pages = {025003},
  doi = {10.1103/revmodphys.95.025003},
  url = {https://doi.org/10.1103/revmodphys.95.025003},
}

@article{Raveendran2023,
  title = {Soft syndrome iterative decoding of quantum {LDPC} codes and hardware architectures},
  author = {Raveendran, Nithin and Valls, Javier and Pradhan, Asit Kumar and Rengaswamy, Narayanan and Garcia-Herrero, Francisco and Vasi{\'{c}}, Bane},
  journal = {EPJ Quantum Technol.},
  year = {2023},
  volume = {10},
  number = {1},
  pages = {45},
  doi = {10.1140/epjqt/s40507-023-00201-1},
  url = {https://doi.org/10.1140/epjqt/s40507-023-00201-1},
}

@article{guerlin2007photon,
  title = {Progressive field-state collapse and quantum non-demolition photon counting},
  author = {Guerlin, Christine and Bernu, Julien and Del{\'{e}}glise, Samuel and Sayrin, Cl{\'{e}}ment and Gleyzes, S{\'{e}}bastien and Kuhr, Stefan and Brune, Michel and Raimond, Jean-Michel and Haroche, Serge},
  journal = {Nature (London)},
  year = {2007},
  volume = {448},
  number = {7156},
  pages = {889-893},
  doi = {10.1038/nature06057},
  url = {https://doi.org/10.1038/nature06057},
}

@article{dasari2022purification,
  title = {Anti-{Zeno} purification of spin baths by quantum probe measurements},
  author = {Dasari, Durga Bhaktavatsala Rao and Yang, Sen and Chakrabarti, Arnab and Finkler, Amit and Kurizki, Gershon and Wrachtrup, J{\"{o}}rg},
  journal = {Nat. Commun.},
  year = {2022},
  volume = {13},
  number = {1},
  pages = {7527},
  doi = {10.1038/s41467-022-35045-3},
  url = {https://doi.org/10.1038/s41467-022-35045-3},
}

@article{bultink2020leakage,
  title = {Protecting quantum entanglement from leakage and qubit errors via repetitive parity measurements},
  author = {Bultink, C. C. and O'Brien, T. E. and Vollmer, R. and Muthusubramanian, N. and Beekman, M. W. and Rol, M. A. and Fu, X. and Tarasinski, B. and Ostroukh, V. and Varbanov, B. and Bruno, A. and DiCarlo, L.},
  journal = {Sci. Adv.},
  year = {2020},
  volume = {6},
  number = {12},
  pages = {eaay3050},
  doi = {10.1126/sciadv.aay3050},
  url = {https://doi.org/10.1126/sciadv.aay3050},
}

@article{xue2020soft,
  title = {Repetitive Quantum Nondemolition Measurement and Soft Decoding of a Silicon Spin Qubit},
  author = {Xue, Xiao and D'Anjou, Benjamin and Watson, Thomas F. and Ward, Daniel R. and Savage, Donald E. and Lagally, Max G. and Friesen, Mark and Coppersmith, Susan N. and Eriksson, Mark A. and Coish, William A. and Vandersypen, Lieven M. K.},
  journal = {Phys. Rev. X},
  year = {2020},
  volume = {10},
  number = {2},
  pages = {021006},
  doi = {10.1103/physrevx.10.021006},
  url = {https://doi.org/10.1103/physrevx.10.021006},
}

@article{kiilerich2016,
 author={Kiilerich, Alexander Holm and M{\o}lmer, Klaus},
 title={{Bayesian} parameter estimation by continuous homodyne detection},
 journal={Phys. Rev. A}, volume={94}, pages={032103}, year={2016}, doi={10.1103/PhysRevA.94.032103}
}

@article{flurin2020,
 author={Flurin, E. and Martin, L. S. and Hacohen-Gourgy, S. and Siddiqi, I.},
 title={Using a Recurrent Neural Network to Reconstruct Quantum Dynamics of a Superconducting Qubit from Physical Observations},
 journal={Phys. Rev. X}, volume={10}, pages={011006}, year={2020}, doi={10.1103/PhysRevX.10.011006}
}

@article{angelatos2021,
 author={Angelatos, Gerasimos and Khan, Saeed A. and T{\"u}reci, Hakan E.},
 title={Reservoir Computing Approach to Quantum State Measurement},
 journal={Phys. Rev. X}, volume={11}, pages={041062}, year={2021}, doi={10.1103/PhysRevX.11.041062}
}

@article{koolstra2022,
 author={Koolstra, G. and Stevenson, N. and Barzili, S. and Burns, L. and Siva, K. and Greenfield, S. and Livingston, W. and Hashim, A. and Naik, R. K. and Kreikebaum, J. M. and O'Brien, K. P. and Santiago, D. I. and Dressel, J. and Siddiqi, I.},
 title={Monitoring Fast Superconducting Qubit Dynamics Using a Neural Network},
 journal={Phys. Rev. X}, volume={12}, pages={031017}, year={2022}, doi={10.1103/PhysRevX.12.031017}
}

@article{dehghani2023,
 author={Dehghani, Hossein and Lavasani, Ali and Hafezi, Mohammad and Gullans, Michael J.},
 title={Neural-network decoders for measurement induced phase transitions},
 journal={Nat. Commun.}, volume={14}, pages={2918}, year={2023}, doi={10.1038/s41467-023-37902-1}
}

@article{Roth2017,
  author = {Roth, Marco and Ganzhorn, Marc and Moll, Nikolaj and Filipp, Stefan and Salis, Gian and Schmidt, Sebastian},
  title = {Analysis of a parametrically driven exchange-type gate and a two-photon excitation gate between superconducting qubits},
  journal = {Phys. Rev. A},
  volume = {96},
  pages = {062323},
  year = {2017},
  doi = {10.1103/PhysRevA.96.062323}
}

@article{Wei2024,
  author = {Wei, Ken Xuan and Lauer, Isaac and Pritchett, Emily and Shanks, William and McKay, David C. and Javadi-Abhari, Ali},
  title = {Native Two-Qubit Gates in Fixed-Coupling, Fixed-Frequency Transmons Beyond Cross-Resonance Interaction},
  journal = {PRX Quantum},
  volume = {5},
  pages = {020338},
  year = {2024},
  doi = {10.1103/PRXQuantum.5.020338}
}

@article{Kotibhaskar2024,
  author = {Kotibhaskar, Nikhil and Shih, Chung-You and Motlakunta, Sainath and Vogliano, Anthony and Hahn, Lewis and Chen, Yu-Ting and Islam, Rajibul},
  title = {Programmable {XY}-type couplings through parallel spin-dependent forces on the same trapped ion motional modes},
  journal = {Phys. Rev. Res.},
  volume = {6},
  pages = {033038},
  year = {2024},
  doi = {10.1103/PhysRevResearch.6.033038}
}

\section*{End Matter}
To understand the sector classification results in the main text, we derive the feature centers and quantify their finite-record fluctuations.

\textit{Stationary feature centers.---}
Repeated probe measurements preserve the occupied sector but can change the state within it. When the outcome-averaged dynamics relaxes to a stationary state $\rho_q^*$, the influence of the initial preparation on the record statistics can decay. These stationary expectations define sector-dependent centers for the first-order and pair features. For our RIM channel, $\rho_q^*=I_{d_q}/d_q$ is stationary, and the examples are prepared in this state from the outset; there is no initial relaxation transient in their ensemble means.

For the outcome maps $\mathcal M_{q,x}(\rho)=M_x^{(q)}\rho M_x^{(q)\dagger}$, define
\begin{equation}
\mathcal E_q=\sum_x\mathcal M_{q,x},\qquad
\mathcal J_q=\sum_x x\mathcal M_{q,x}.
\label{eq:em_maps}
\end{equation}
The channel $\mathcal E_q$ describes a cycle when its outcome is ignored. The map $\mathcal J_q$ instead weights each outcome by its recorded value. The steady outcome mean is therefore
\begin{equation}
m_q=\sum_{x=\pm1}x\,\Tr[\mathcal M_{q,x}(\rho_q^*)]
=\Tr[\mathcal J_q(\rho_q^*)].
\label{eq:em_mean}
\end{equation}
For a pair separated by $k$ cycles, the first measurement conditions the system state, the outcomes of the $k-1$ intermediate cycles are averaged over, and the second measurement probes the resulting state. Following this sequence gives
\begin{equation}
C_q(k)=\Tr[\mathcal J_q\mathcal E_q^{k-1}\mathcal J_q(\rho_q^*)],
\quad k\ge1,
\label{eq:em_centers}
\end{equation}
This expression is exact and does not require weak measurements~\cite{Jin2026noise}. With the stationary preparations used here, the outcome-averaged evolution preserves $\rho_q^*$. Every cycle therefore has the same outcome mean, and every pair at a fixed lag has the same correlation. Averaging over the $L$ outcomes or the $L-k$ pairs leaves these ensemble means unchanged:
\begin{equation}
\mathbb E_q[S_0]=m_q,\qquad \mathbb E_q[S_k]=C_q(k).
\label{eq:em_feature_means}
\end{equation}
These relations hold for every allowed $L$, although individual records still fluctuate. A nonstationary preparation can introduce an initial transient, whose contribution to a long-record average becomes small if relaxation occurs well within the observation window. If the dynamics can remain trapped in smaller subspaces within a sector, the long-record statistics may also depend on the initial preparation.

The centers can differ between sectors because the probe couples differently to each sector and samples different internal dynamics. The single-outcome center $m_q$ measures the steady probe response in sector $q$. Pair centers additionally depend on how the first measurement conditions the system state and how these changes affect the later readout through $\mathcal E_q^{k-1}$. Even sectors with the same form of maximally mixed steady state and identical $m_q$ can have different $C_q(k)$, as in the parity example. Conservation alone does not guarantee distinct centers; their separation depends on the instrument and on the lags retained in $F_K$.

The sector dependence of pair correlations can also be understood from a spectral perspective. A diagonalizable channel has the spectral decomposition
\begin{equation}
\widehat{\mathcal E}_q=\sum_\alpha\lambda_{q,\alpha}
\dket{R_{q,\alpha}}\dbra{L_{q,\alpha}}.
\label{eq:em_channel_spectral}
\end{equation}
Here hats denote matrices acting on vectorized operators, with $\dbra{A}\dket{B}=\Tr(A^\dagger B)$. The eigenvalue $\lambda_{q,\alpha}$ describes mode $\alpha$, whose right and left eigenvectors are $\dket{R_{q,\alpha}}$ and $\dbra{L_{q,\alpha}}$, normalized by $\dbra{L_{q,\alpha}}\dket{R_{q,\beta}}=\delta_{\alpha\beta}$. Equation~\eqref{eq:em_centers} then gives
\begin{equation}
C_q(k)=\dbra{I_q}\widehat{\mathcal J}_q
\widehat{\mathcal E}_q^{k-1}\widehat{\mathcal J}_q\dket{\rho_q^*}
=\sum_\alpha d_{q,\alpha}\lambda_{q,\alpha}^{k-1},
\label{eq:em_spectral}
\end{equation}
where $d_{q,\alpha}=\dbra{I_q}\widehat{\mathcal J}_q\dket{R_{q,\alpha}}\dbra{L_{q,\alpha}}\widehat{\mathcal J}_q\dket{\rho_q^*}$, with $I_q$ the identity within sector $q$. These weights specify which dynamical modes contribute to the measured correlation. For a unique steady state, its mode can be chosen as $\lambda_{q,0}=1$, $\dket{R_{q,0}}=\dket{\rho_q^*}$, and $\dbra{L_{q,0}}=\dbra{I_q}$. Its weight is therefore $d_{q,0}=\Tr[\mathcal J_q(\rho_q^*)]^2=m_q^2$, leaving
\begin{equation}
C_q(k)-m_q^2=\sum_{\alpha\ne0}d_{q,\alpha}\lambda_{q,\alpha}^{k-1}.
\label{eq:em_connected_modes}
\end{equation}
If all remaining modes have $|\lambda_{q,\alpha}|<1$, the right-hand side vanishes as $k$ increases, giving $C_q(k)\to m_q^2$. At finite lag these terms can remain appreciable. Two sectors with identical $m_q$ can then have different pair centers because their mode weights or eigenvalues differ. This is the additional sector information captured by finite-lag features. A mode with $|\lambda_{q,\alpha}|=1$ and $\lambda_{q,\alpha}\ne1$ instead contributes an undamped oscillation when $d_{q,\alpha}\ne0$. In this case, $C_q(k)$ need not approach $m_q^2$ as the lag grows, even for a stationary preparation. When the oscillation frequencies or weights differ between sectors, suitable lags can provide additional sector information and improve classification.

\textit{Finite-record fluctuations.---}
For a finite measurement record, the measured features fluctuate around the centers derived above. We show that, when the correlations between feature samples decay sufficiently rapidly, the width of a pair-feature distribution at fixed lag $k$ typically decreases as $(L-k)^{-1/2}$ for long records. Write $S_k=N^{-1}\sum_{n=1}^N Z_n^{(k)}$, where $Z_n^{(0)}=x_n$ and $N=L$ for $S_0$, or $Z_n^{(k)}=x_nx_{n+k}$ and $N=L-k$ for a pair feature. For the stationary preparations used here, define $\gamma_{q,k}(t)=\operatorname{Cov}_q(Z_n^{(k)},Z_{n+t}^{(k)})$. The exact variance is
\begin{equation}
\operatorname{Var}_q(S_k)=\frac1N\left[\gamma_{q,k}(0)
+2\sum_{t=1}^{N-1}\left(1-\frac tN\right)\gamma_{q,k}(t)\right].
\label{eq:em_variance}
\end{equation}
The variance expansion contains $N$ single-sample contributions, each equal to $\gamma_{q,k}(0)$, and $N-t$ sample pairs at each separation $t\ge1$. Counting both orderings of each pair gives the factor $2(N-t)$. If the feature samples $Z_n^{(k)}$ are independent, all terms with $t>0$ vanish and the variance reduces to $\gamma_{q,k}(0)/N$; correlations between these samples contribute the additional covariance terms.

If the memory between samples fades sufficiently quickly, the bracket approaches a record-length-independent value once $N$ greatly exceeds the memory time. The feature width then decreases as $N^{-1/2}$, giving the $(L-k)^{-1/2}$ dependence for pair features and $L^{-1/2}$ for the outcome mean. Slowly decaying correlations can delay this scaling, while persistent correlations can prevent it. When these widths shrink, distinct sector-center vectors $(m_q,C_q(1),\ldots,C_q(K))$ become easier to distinguish.

\end{document}

\title{Supplemental Material: Distinguishing Conserved Quantum Sectors via Measurement Record Features}
\author{Hao Zhu}
\affiliation{State Key Laboratory of Semiconductor Physics and Chip Technologies, Institute of Semiconductors, Chinese Academy of Sciences, Beijing 100083, China}
\author{Guo-Feng Zhang}
\email{gf1978zhang@buaa.edu.cn}
\affiliation{School of Physics, Beihang University, Beijing 100191, China}
\author{Wen-Long Ma}
\email{wenlongma@semi.ac.cn}
\affiliation{State Key Laboratory of Semiconductor Physics and Chip Technologies, Institute of Semiconductors, Chinese Academy of Sciences, Beijing 100083, China}
\affiliation{Center of Materials Science and Opto-Electronic Technology, University of Chinese Academy of Sciences, Beijing 100049, China}
\date{\today}
\maketitle
\tableofcontents

\section{Conserved sectors and repeated measurements}\label{sec:conserved}

\subsection{Conservation laws in quantum channels}

In a closed quantum system governed by a time-independent
Hamiltonian $H_S$, an observable $Q$ is conserved if
\begin{align}
[H_S,Q]=0.
\end{align}
This condition ensures that the expectation value of $Q$
remains unchanged during the unitary evolution for any
initial state.
For a quantum system whose evolution is described by a
quantum channel $\mathcal E$, the notion of conservation
can be formulated in terms of the adjoint channel
$\mathcal E^\dagger$.
An observable $Q$ is conserved if
\begin{align}
\mathcal E^\dagger(Q)=Q,
\end{align}
which ensures that its expectation value remains unchanged
under the channel for any state $\rho$,
\begin{align}
\Tr[Q\mathcal E(\rho)]
&=\Tr[\mathcal E^\dagger(Q)\rho]
=\Tr(Q\rho).
\end{align}
In this work, we focus on the conservation of a complete set of orthogonal projectors $\{\Pi_q\}$, which define a decomposition of the system Hilbert space into conserved sectors.

We first consider the outcome maps and the corresponding quantum channel,
\begin{align}
\mathcal M_x(\rho)&=M_x\rho M_x^\dagger,\quad
\mathcal E(\rho)=\sum_x\mathcal M_x(\rho).
\end{align}
Let $\{\Pi_q\}$ be a complete set of orthogonal projectors satisfying
\begin{align}
\Pi_q\Pi_r=\delta_{qr}\Pi_q,\qquad
\sum_q\Pi_q=I.
\end{align}
We assume that every Kraus operator preserves each sector, i.e.,
\begin{align}
[M_x,\Pi_q]=0
\end{align}
for all $x$ and $q$.
For a measurement record $\bm x=(x_1,\ldots,x_L)$,
the corresponding operator
\begin{align}
M_{\bm x}=M_{x_L}\cdots M_{x_1}
\end{align}
also commutes with every sector projector,
\begin{align}
[M_{\bm x},\Pi_q]=0,
\end{align}
and is therefore block diagonal with respect to the same sector decomposition.
Inserting the resolution of the identity on both sides of the initial state $\rho_0$, we obtain
\begin{align}
P^{(L)}(\bm x)
=\Tr(M_{\bm x}\rho_0M_{\bm x}^\dagger)
=\sum_{q,r}\Tr(M_{\bm x}\Pi_q\rho_0\Pi_rM_{\bm x}^\dagger).
\end{align}
Since $M_{\bm x}$ commutes with every $\Pi_q$,
the off-diagonal contributions vanish for $q\ne r$,
\begin{align}
&\Tr(M_{\bm x}\Pi_q\rho_0\Pi_rM_{\bm x}^\dagger)
=\Tr(\Pi_r\Pi_qM_{\bm x}^\dagger M_{\bm x}\rho_0)
=0.
\end{align}
Defining the weight $\mu_q$ and the normalized initial state
$\eta_q$ in sector $q$ as
\begin{align}
\mu_q=\Tr(\Pi_q\rho_0),\qquad
\eta_q=\frac{\Pi_q\rho_0\Pi_q}{\mu_q},
\end{align}
where $\eta_q$ is defined for $\mu_q>0$, we obtain
\begin{align}
P^{(L)}(\bm x)=\sum_{q:\mu_q>0}\mu_qP_q^{(L)}(\bm x),
\end{align}
with the sector-dependent record probability given by
\begin{align}
P_q^{(L)}(\bm x)
=\Tr(M_{\bm x}\eta_qM_{\bm x}^\dagger).
\end{align}
Therefore, initial coherences between different sectors do not contribute to the measurement record probabilities.
The record statistics are identical to those obtained from the incoherent mixture $\sum_q\Pi_q\rho_0\Pi_q$.

We next establish the equivalence between the conservation of a sector projector by the quantum channel and its conservation by every Kraus operator.
The adjoint channel is defined as
\begin{align}
\mathcal E^\dagger(O)=\sum_xM_x^\dagger O M_x.
\end{align}
If $[M_x,\Pi_q]=0$ for all $x$, the completeness relation $\sum_xM_x^\dagger M_x=I$ immediately gives
\begin{align}
\mathcal E^\dagger(\Pi_q)
=\sum_xM_x^\dagger\Pi_qM_x
=\Pi_q\sum_xM_x^\dagger M_x
=\Pi_q.
\end{align}
Conversely, suppose that $\mathcal E^\dagger(\Pi_q)=\Pi_q$.
For any vector $\ket{v}$ in sector $q$, satisfying
$\Pi_q\ket{v}=\ket{v}$, we have
\begin{align}
\sum_x\|(I-\Pi_q)M_x\ket{v}\|^2
=\sum_x\bra{v}M_x^\dagger(I-\Pi_q)M_x\ket{v}
=\bra{v}I-\mathcal E^\dagger(\Pi_q)\ket{v}
=\bra{v}(I-\Pi_q)\ket{v}
=0.
\end{align}
Since every term in the sum is nonnegative, it follows that
\begin{align}
(I-\Pi_q)M_x\Pi_q=0
\end{align}
for every $x$.
Similarly, for any vector $\ket{w}$ in the orthogonal complement of sector $q$, satisfying $\Pi_q\ket{w}=0$, we have
\begin{align}
\sum_x\|\Pi_qM_x\ket{w}\|^2
=\bra{w}\mathcal E^\dagger(\Pi_q)\ket{w}
=\bra{w}\Pi_q\ket{w}
=0.
\end{align}
This implies that
\begin{align}
\Pi_qM_x(I-\Pi_q)=0
\end{align}
for every $x$.
Both off-diagonal blocks of $M_x$ therefore vanish, yielding
\begin{align}
M_x\Pi_q=\Pi_qM_x\Pi_q=\Pi_qM_x.
\end{align}

We have thus established the equivalence
\begin{align}
\mathcal E^\dagger(\Pi_q)=\Pi_q
\quad\Longleftrightarrow\quad
[M_x,\Pi_q]=0\quad\forall x.
\end{align}
Consequently, a state initially supported in sector $q$ remains in that sector under both the unconditional evolution and the evolution conditioned on individual measurement outcomes.

\subsection{Ramsey interferometric measurements}

We now derive the Kraus operators for the Ramsey interferometric measurement (RIM) protocol introduced in the main text.
Setting $\hbar=1$, the joint evolution of the probe and the system during the interaction interval $T$ is given by 
\begin{align}
U_{aS}
&=e^{-i(I_a\otimes H_S+\sigma_a^z\otimes B)T}
=\ket{0}_a\bra{0}\otimes U_+
+\ket{1}_a\bra{1}\otimes U_-,
\end{align}
where
\begin{align}
U_\pm=e^{-i(H_S\pm B)T}.
\end{align}
The two $\pi/2$ pulses applied to the probe are described by
\begin{align}
R_\theta
&=\exp\left[-\frac{i\pi}{4}
(\cos\theta\,\sigma_a^x+\sin\theta\,\sigma_a^y)\right]
=\frac{1}{\sqrt{2}}
\begin{pmatrix}
1&-ie^{-i\theta}\\
-ie^{i\theta}&1
\end{pmatrix}.
\end{align}
For an initial joint state $\ket{0}_a\otimes\ket{\psi}$,
the first pulse $R_{\phi_1}$ prepares the state
\begin{align}
\frac{1}{\sqrt{2}}
\left(\ket{0}_a-ie^{i\phi_1}\ket{1}_a\right)
\otimes\ket{\psi}.
\end{align}
After the interaction interval $T$, the joint state becomes
\begin{align}
\frac{1}{\sqrt{2}}
\left(
\ket{0}_a\otimes U_+\ket{\psi}
-ie^{i\phi_1}\ket{1}_a\otimes U_-\ket{\psi}
\right).
\end{align}
Applying the second pulse $R_{\phi_2}$ gives
\begin{align}
\ket{\Psi_{\mathrm f}}
=\ket{0}_a\otimes A_0\ket{\psi}
+\ket{1}_a\otimes A_1\ket{\psi},
\end{align}
where
\begin{align}
A_0&=\frac12(U_+-e^{i\phi}U_-),\qquad
A_1=-\frac{ie^{i\phi_2}}2(U_++e^{i\phi}U_-),
\end{align}
and $\phi=\phi_1-\phi_2$ is the relative phase between
the two pulses.

We label the projective measurement outcomes of the probe by $\alpha\in\{0,1\}$ and associate each outcome with a binary variable
\begin{align}
x=(-1)^\alpha,
\end{align}
so that $x=+1$ corresponds to $\ket{0}_a$ and $x=-1$ corresponds to $\ket{1}_a$.
An outcome-dependent overall phase of a Kraus operator has no effect on the corresponding outcome map, since
\begin{align}
(e^{i\chi_x}M_x)\rho(e^{i\chi_x}M_x)^\dagger
=M_x\rho M_x^\dagger.
\end{align}
Omitting such phases, we obtain the Kraus operators used in the main text,
\begin{align}
M_x=\frac12(U_+-xe^{i\phi}U_-),\qquad x=\pm1.
\label{eq:ramsey}
\end{align}

Summing over the two measurement outcomes, the quantum
channel associated with one RIM cycle is given by
\begin{align}
\mathcal E(\rho)
&=\sum_{x=\pm1}M_x\rho M_x^\dagger
=\frac12\left(
U_+\rho U_+^\dagger+U_-\rho U_-^\dagger
\right).
\end{align}
The unitarity of $U_\pm$ implies that the Kraus operators
satisfy both
\begin{align}
\sum_xM_x^\dagger M_x=I,\qquad
\sum_xM_xM_x^\dagger=I.
\end{align}
The first relation ensures that the channel is trace preserving,
while the second ensures that it is unital.

If both $H_S$ and $B$ commute with a sector projector $\Pi_q$,
\begin{align}
[H_S,\Pi_q]=[B,\Pi_q]=0,
\end{align}
then the conditional evolution operators $U_\pm$ and
the resulting Kraus operators $M_x$ also commute with $\Pi_q$.
Let $M_x^{(q)}$ and $U_\pm^{(q)}$ denote the restrictions
of these operators to sector $q$.
Within this sector, the quantum channel takes the form
\begin{align}
\mathcal E_q(\rho)
&=\sum_xM_x^{(q)}\rho M_x^{(q)\dagger}
=\frac12\sum_{s=\pm}
U_s^{(q)}\rho U_s^{(q)\dagger}.
\end{align}
Since $U_\pm^{(q)}$ are unitary within the sector,
$\mathcal E_q$ is also unital.
Consequently, the maximally mixed state
\begin{align}
\eta_q^{\mathrm{mm}}=\frac{I_{d_q}}{d_q}
\end{align}
is a stationary state of $\mathcal E_q$,
\begin{align}
\mathcal E_q(\eta_q^{\mathrm{mm}})=\eta_q^{\mathrm{mm}},
\end{align}
where $d_q=\Tr(\Pi_q)$ is the dimension of sector $q$.

\subsection{Numerical record generation}

To numerically generate a measurement record, we sequentially
sample the measurement outcomes and update the system state.
Let $\rho_{n-1}$ be the normalized system state conditioned
on the previous outcomes $(x_1,\ldots,x_{n-1})$.

At the $n$th measurement cycle, we sample an outcome
$x_n\in\{+1,-1\}$ according to the conditional probability
\begin{align}
p(x_n|x_1,\ldots,x_{n-1})
&=\Tr[\mathcal M_{x_n}(\rho_{n-1})]
=\Tr(M_{x_n}\rho_{n-1}M_{x_n}^\dagger).
\end{align}
After obtaining $x_n$, the system state is updated according to
\begin{align}
\rho_n=
\frac{M_{x_n}\rho_{n-1}M_{x_n}^\dagger}
{p(x_n|x_1,\ldots,x_{n-1})}.
\end{align}
Repeating this procedure for $L$ measurement cycles generates
a record $\bm x=(x_1,\ldots,x_L)$.

For numerical simulations, propagating a pure state is
computationally less expensive than propagating a density matrix.
We therefore represent the maximally mixed state within
sector $q$ as a uniform ensemble of orthonormal basis states,
\begin{align}
\eta_q^{\mathrm{mm}}
=\frac{I_{d_q}}{d_q}
=\frac{1}{d_q}\sum_{j=1}^{d_q}\ket{q,j}\bra{q,j},
\end{align}
where $\{\ket{q,j}\}_{j=1}^{d_q}$ is an orthonormal basis
of the sector.
To generate each trajectory, we randomly select an initial pure
state $\ket{q,j}$ with probability $1/d_q$ and propagate
the corresponding state vector.
For the spin models whose sectors are spanned by computational
basis states, we directly use these states as the sampling basis.

Given the pure state $\ket{\psi_{n-1}}$ before the
$n$th measurement, the probability of outcome $x_n$ is
\begin{align}
p(x_n|x_1,\ldots,x_{n-1};j)
=\|M_{x_n}^{(q)}\ket{\psi_{n-1}}\|^2,
\end{align}
and the state is updated according to
\begin{align}
\ket{\psi_n}
=\frac{M_{x_n}^{(q)}\ket{\psi_{n-1}}}
{\sqrt{p(x_n|x_1,\ldots,x_{n-1};j)}}.
\end{align}
This pure-state sampling procedure reproduces exactly the
measurement record distribution obtained from the maximally
mixed initial state.
To see this, consider the probability of a given record
$\bm x$, averaged over the uniformly sampled initial pure states,
\begin{align}
P_q^{(L)}(\bm x)
&=\frac{1}{d_q}\sum_{j=1}^{d_q}
\|M_{\bm x}^{(q)}\ket{q,j}\|^2
=\frac{1}{d_q}\sum_{j=1}^{d_q}
\Tr\left(
M_{\bm x}^{(q)}\ket{q,j}\bra{q,j}
M_{\bm x}^{(q)\dagger}
\right)
=\Tr\left(
M_{\bm x}^{(q)}\eta_q^{\mathrm{mm}}
M_{\bm x}^{(q)\dagger}
\right),
\end{align}
where $M_{\bm x}^{(q)}=M_{x_L}^{(q)}\cdots M_{x_1}^{(q)}$.
Thus, the pure-state sampling procedure generates records
from the same probability distribution as direct simulations
starting from the maximally mixed density matrix,
without requiring density-matrix propagation.

\section{Classification framework}\label{sec:methods}

\subsection{Record features}\label{subsec:features}

To classify measurement records without directly evaluating
their full probability distributions, we extract a set of
statistical features from each record
$\bm x=(x_1,\ldots,x_L)$, where $x_n\in\{+1,-1\}$.
We first consider the single-outcome mean and the
two-outcome correlations,
\begin{align}
S_0(\bm x)&=\frac{1}{L}\sum_{n=1}^{L}x_n,\qquad
S_k(\bm x)=\frac{1}{L-k}
\sum_{n=1}^{L-k}x_nx_{n+k},
\quad 1\leq k<L.
\end{align}
Combining these statistics, we define the feature vector
\begin{align}
F_K(\bm x)=(S_0,S_1,\ldots,S_K),
\label{eq:FK}
\end{align}
where $0\leq K<L$.
Here $K$ specifies the maximum lag included in the feature vector.
The statistic $S_0$ is of first order, whereas each $S_k$
with $k\geq1$ is of second order.
Thus, $F_0$ contains only the single-outcome mean, while
$F_K$ with $K\geq1$ additionally contains two-outcome
correlations at lags from 1 to $K$.
Increasing $K$ extends the range of temporal correlations
without increasing their order.
All components of $F_K$ can be evaluated in $O[L(K+1)]$
operations.
By maintaining running sums and storing only the most
recent $K$ outcomes, the feature vector can also be
computed sequentially with $O(K+1)$ memory, without
storing the entire measurement record.

For numerical fitting, it is convenient to represent
these statistics using integer counts,
\begin{align}
N_+&=\sum_{n=1}^{L}\frac{1+x_n}{2}
=\frac{L(1+S_0)}{2},\qquad
N_k=\sum_{n=1}^{L-k}\frac{1+x_nx_{n+k}}{2}
=\frac{(L-k)(1+S_k)}{2}.
\end{align}
Here $N_+$ counts the number of positive outcomes, and
$N_k$ counts the number of equal-outcome pairs separated
by lag $k$.
At fixed $L$, the transformations between these counts
and the corresponding statistics are invertible.
Thus, the two representations contain exactly the
same information.

More generally, we can construct statistical features
involving products of an arbitrary number of
measurement outcomes.
Let $\mathcal A$ be a nonempty set of nonnegative integer
offsets satisfying
\begin{align}
\min\mathcal A=0,\qquad b=\max\mathcal A<L.
\end{align}
We define the corresponding statistic as
\begin{align}
C_{\mathcal A}^{(L)}(\bm x)
=\frac{1}{L-b}
\sum_{n=1}^{L-b}\prod_{a\in\mathcal A}x_{n+a}.
\label{eq:generalproducts}
\end{align}
The previously defined statistics are recovered as
\begin{align}
S_0=C_{\{0\}}^{(L)}(\bm x),\qquad
S_k=C_{\{0,k\}}^{(L)}(\bm x).
\end{align}

For each starting position $n$, the product involves the
measurement outcomes at positions $n+a$ for all
$a\in\mathcal A$.
The largest separation between any two of these outcomes
is $b$ measurement cycles.
The sum therefore contains $L-b$ terms, ensuring that
all outcomes lie within the record.
For example, choosing $\mathcal A=\{0,1,3\}$ gives
the third-order statistic
\begin{align}
C_{\{0,1,3\}}^{(L)}(\bm x)
=\frac{1}{L-3}
\sum_{n=1}^{L-3}x_nx_{n+1}x_{n+3}.
\end{align}
The product order $m=|\mathcal A|$ specifies the number
of outcomes involved in each product, while the maximum
lag $b$ specifies its temporal range.
These are two distinct characteristics of a statistical
feature, subject to the constraint $m\leq b+1$.

The notation used for statistical features up to fifth
order is summarized in the following table, where the
nonzero offsets in each row are strictly increasing.

\begin{center}
\begin{tabular}{lccc}
\toprule
Offset set $\mathcal A$ & Notation & Order & Maximum lag\\
\midrule
$\{0\}$ & $S_0$ & 1 & 0\\
$\{0,k\}$ & $S_k$ & 2 & $k$\\
$\{0,a,b\}$ & $T_{a,b}$ & 3 & $b$\\
$\{0,a,b,c\}$ & $U_{a,b,c}$ & 4 & $c$\\
$\{0,a,b,c,d\}$ & $V_{a,b,c,d}$ & 5 & $d$\\
\bottomrule
\end{tabular}
\end{center}

This construction provides a systematic way to extend
the feature vectors beyond second-order correlations.
In particular, higher-order statistics can be combined
with $F_K$ to construct additional classifiers for the
many-spin systems in Examples III and IV.

\subsection{Feature-based classifiers}

Using the record features, we now describe how to construct and implement feature-based classifiers.
For a given feature map $F$, the probability of observing a feature value $f$ in sector $q$ is
\begin{align}
p_{q,F}^{(L)}(f)
=\sum_{\bm x:F(\bm x)=f}P_q^{(L)}(\bm x).
\end{align}
In numerical implementations, the exact feature distributions $p_{q,F}^{(L)}$ are generally not available.
We therefore collect training records of length $L$ from each known sector $q$ and use their feature values to estimate the corresponding distributions $\widehat p_{q,F}^{(L)}(f)$.
For a new measurement record, the classifier evaluates $F(\bm x)$ and obtains its estimated probability in each sector.

Each probability is weighted by the initial sector weight $\mu_q$, and the record is assigned to the sector with the largest weighted probability,
\begin{align}
\widehat Q_F(\bm x)
=\arg\max_q\left[
\mu_q\widehat p_{q,F}^{(L)}(F(\bm x))
\right].
\end{align}
Depending on the dimension of the feature vector, we use either histogram-based estimates or multivariate Gaussian models.

For the integer count representation introduced above, we estimate the feature distributions by constructing histograms from the training records in each sector.
To reduce the effect of empty bins arising from finite training data, we add a pseudocount of $0.25$ to each bin of a scalar histogram.
To reduce fluctuations in the estimated distributions, we optionally smooth the histograms with a Gaussian filter using zero padding at the boundaries.
For scalar histograms, the standard deviation of the Gaussian filter, measured in units of histogram bins, is selected from $\sigma\in\{0,0.75,1.5,3\}$ according to the classification infidelity on the validation set, where $\sigma=0$ corresponds to no smoothing.
Each sector-conditioned histogram is normalized separately such that $\sum_f\widehat p_{q,F}^{(L)}(f)=1$.

As the dimension of the feature vector increases, estimating the joint feature distribution using multidimensional histograms becomes increasingly difficult.
We therefore also use multivariate Gaussian models to approximate these distributions.
Each feature coordinate is standardized using the mean and standard deviation calculated from the training data.
For each sector $q$, we estimate the mean vector $\widehat{\bm m}_q$ and covariance matrix $\widehat\Sigma_q$ of the standardized features.
To improve numerical stability, we use the regularized covariance matrix $\Sigma_{q,r}=(1-r)\widehat\Sigma_q+rI$, with $r\in\{0.001,0.01,0.05,0.2,0.5\}$ selected using the validation set.
The logarithm of the weighted Gaussian density, up to a sector-independent constant, gives the classification score
\begin{align}
g_q(\widetilde{\bm f})
={}&\ln\mu_q-\frac12\ln\det\Sigma_{q,r}
-\frac12(\widetilde{\bm f}-\widehat{\bm m}_q)^T
\Sigma_{q,r}^{-1}
(\widetilde{\bm f}-\widehat{\bm m}_q),
\end{align}
where $\widetilde{\bm f}$ is the standardized feature vector associated with $F(\bm x)$.
The corresponding classifier is given by $\widehat Q_F(\bm x)=\arg\max_q g_q(\widetilde{\bm f})$.

For the numerical implementation, the measurement records are divided into three disjoint data sets for training, validation, and testing.
The training set is used to estimate feature distributions or fit classifier parameters, the validation set is used for model and hyperparameter selection, and the test set is used exclusively to evaluate the classification fidelity of the selected classifiers.
To evaluate classification fidelity at different record lengths $L$, we reuse prefixes of the same measurement trajectories within each data set.
The division is performed at the trajectory level, ensuring that prefixes of a given trajectory never appear in different data sets.

The performance of a feature-based classifier depends on both the record length and the choice of features.
For stationary and ergodic measurement records within sector $q$, the feature vector defined in Eq.~\eqref{eq:FK} at fixed $K$ converges to
\begin{align}
F_K(\bm x)
\xrightarrow{L\to\infty}
\big(m_q,C_q(1),\ldots,C_q(K)\big),
\end{align}
where $m_q=\mathbb E_q[x_n]$ and $C_q(k)=\mathbb E_q[x_nx_{n+k}]$ are the stationary outcome mean and the stationary correlation at lag $k$, respectively.
If these limiting vectors are distinct across sectors, the feature distributions concentrate around different values as $L$ increases, enabling reliable sector identification.
Explicit channel expressions for these limiting values, the conditions for convergence, and finite-record fluctuations are discussed in the End Matter of the main text.

At a fixed record length $L$, the effect of including correlations at additional lags can be understood from the nested structure of the feature vectors $F_K$.
Let $\mathcal F_F^*(L)$ denote the optimal classification fidelity achievable using the feature $F$, and let $\mathcal F_{\rm full}^*(L)$ denote that achievable using the full measurement record.
Since $F_{K+1}$ contains all the coordinates of $F_K$, any classifier based on $F_K$ can also be implemented using $F_{K+1}$.
Therefore,
\begin{align}
\mathcal F_{F_0}^*(L)
\leq\mathcal F_{F_1}^*(L)
\leq\cdots
\leq\mathcal F_{F_K}^*(L)
\leq\mathcal F_{\rm full}^*(L).
\label{eq:optfid}
\end{align}
A similar hierarchy holds when higher-order statistics are added to the feature vector.
In practice, however, including more features may increase the difficulty of estimating their joint distributions from finite training data and lead to a decrease in the achieved classification fidelity.
For this reason, we retain classifiers based on simpler features as candidates when constructing classifiers with larger feature vectors.

\subsection{Neural-network classifiers}

To compare the performance of feature-based classifiers with that of a classifier using the full measurement record, we consider a feedforward neural network (NN) following the cumulative-count input representation and shallow architecture of Liu et al.~\cite{Liu2020}.

For a measurement record $\bm x=(x_1,\ldots,x_L)$, we construct the input vector $\bm z=(z_1,\ldots,z_L)$, where $z_\ell=\sum_{n=1}^{\ell}(x_n+1)/2$ counts the number of positive measurement outcomes in the first $\ell$ measurement cycles.
Unlike the total count $N_+$ introduced above, the cumulative-count vector retains the full measurement record, since the original outcomes can be recovered as $x_\ell=2(z_\ell-z_{\ell-1})-1$, with $z_0=0$.
Each input coordinate is standardized according to $\widetilde z_\ell=(z_\ell-\overline z_\ell)/\sigma_\ell$, where $\overline z_\ell$ and $\sigma_\ell$ are the mean and standard deviation of the $\ell$th input coordinate calculated from the training data.
The same transformation is applied to the validation and test data.

The NN consists of an input layer, a single hidden layer, and an output layer.
The hidden-layer activations and output scores are given by
\begin{align}
\bm h=\operatorname{sigmoid}(W^T\widetilde{\bm z}+\bm b),
\qquad
s_{q,L}(\bm x)=\operatorname{softmax}_q(V^T\bm h+\bm c),
\end{align}
where $W$ and $V$ are the weight matrices, and $\bm b$ and $\bm c$ are the bias vectors.
We use 32 hidden units with sigmoid activation functions, while the number of output units equals the number of sectors to be classified.
The softmax outputs satisfy $\sum_q s_{q,L}(\bm x)=1$.

The NN is trained using equal numbers of training records from each sector by minimizing the cross-entropy loss with an $L_2$ regularization term,
\begin{align}
\mathcal L
=-\frac{1}{N_{\rm train}}\sum_{i=1}^{N_{\rm train}}
\ln s_{q_i,L}(\bm x_i)
+\frac{\lambda}{2N_{\rm train}}
\left(\|W\|_F^2+\|V\|_F^2\right),
\end{align}
where $N_{\rm train}$ is the total number of training records, $q_i$ is the sector label of the $i$th record, $\|\cdot\|_F$ denotes the Frobenius norm, and $\lambda=10^{-3}$ is the regularization parameter.
We use full-batch L-BFGS-B optimization with at most 250 iterations.
For each record length $L$, we train three NNs with independent random initializations and select the network parameters that minimize the validation cross-entropy loss.

Since the NN is trained on balanced data, we incorporate the initial sector weights $\mu_q$ when classifying a new measurement record.
The resulting classifier is
\begin{align}
\widehat Q_{\rm NN}(\bm x)
=\arg\max_q\left[\mu_q s_{q,L}(\bm x)\right].
\end{align}
The NN provides a numerical benchmark for classification using the full measurement record.
Although its input retains all the information in the record, its classification performance is limited by the network architecture, the amount of training data, and the optimization procedure.
Therefore, such an NN does not necessarily achieve a higher classification fidelity than feature-based classifiers.

\subsection{Classification fidelity}

To evaluate the performance of a trained classifier $C$, we generate $N_q$ independent test records from each sector $q$.
Let $n_q$ denote the number of correctly classified records in sector $q$, so that $\widehat f_q=n_q/N_q$ is the estimated fraction of correct classifications within that sector.
The classification fidelity is estimated by weighting these fractions with the initial sector weights $\mu_q$.

Since the correctness of each classification is a Bernoulli random variable, the variance of the estimated fidelity can be obtained from the independent test samples in each sector.
The estimated classification fidelity and its standard error (SE) are given by
\begin{align}
\widehat{\mathcal F}_C=\sum_q\mu_q\widehat f_q,\qquad
\operatorname{SE}(\widehat{\mathcal F}_C)
=\left[\sum_q\frac{\mu_q^2\widehat f_q(1-\widehat f_q)}{N_q}
\right]^{1/2}.
\end{align}
The classification infidelity $1-\widehat{\mathcal F}_C$ has the same standard error.
The error bars in the main text correspond to $\pm1.96\operatorname{SE}$, giving approximate 95\% confidence intervals for fixed trained classifiers.
When comparing classifiers evaluated on the same test records, we use paired differences in their correctness indicators.

\section{Examples: parameters, data, and readout results}\label{sec:examples}

\subsection{Example I: QND outcome-count readout}
\label{sec:exampleI}

\subsubsection{Outcome-count statistics}

We consider a set of commuting normal Kraus operators with a common orthonormal eigenbasis, satisfying $M_x\ket{j}=c_{xj}\ket{j}$ for all $x$.
Associated with each eigenstate $\ket{j}$ is an eigenvalue vector $\bm c_j=(c_{xj})_x$.
Following the sector definition in the main text, we group eigenstates whose eigenvalue vectors differ only by an overall phase into the same sector.
For any eigenstate $\ket{j}$ within sector $q$, the measurement outcome probabilities are given by $|c_{xj}|^2=(1+xm_q)/2$, where $m_q$ is the mean outcome in that sector.
These probabilities are independent of the state within the sector.
Since each measurement also preserves the sector, successive measurement outcomes are independent and identically distributed within a given sector.

A concrete realization of this QND measurement is provided by the Ramsey protocol with $H_S=0$ and $B=\sum_qb_q\Pi_q$.
Using Eq.~\eqref{eq:ramsey}, the outcome probabilities within sector $q$ are
\begin{align}
p_q(x)
=\frac14\left|e^{-ib_qT}-xe^{i\phi}e^{ib_qT}\right|^2
=\frac{1-x\cos(2b_qT+\phi)}{2},
\end{align}
which gives $m_q=-\cos(2b_qT+\phi)$.
Since the measurement outcomes are independent within each sector, the probability of a measurement record factorizes as
\begin{align}
P_q^{(L)}(\bm x)
=\prod_{n=1}^L p_q(x_n)
=\left(\frac{1+m_q}{2}\right)^{N_+}
\left(\frac{1-m_q}{2}\right)^{L-N_+},
\end{align}
where $N_+=L(1+S_0)/2$ is the number of positive measurement outcomes.

The positive-outcome count follows a binomial distribution, $N_+\sim\operatorname{Binomial}[L,(1+m_q)/2]$.
Using $S_0=2N_+/L-1$, we obtain the exact finite-record distribution
\begin{align}
\Pr_q\left(S_0=\frac{2j}{L}-1\right)
=\binom Lj
\left(\frac{1+m_q}{2}\right)^j
\left(\frac{1-m_q}{2}\right)^{L-j},
\label{eq:qnd_S0_law}
\end{align}
where $j=0,\ldots,L$.
The mean and variance of $S_0$ are
\begin{align}
\mathbb E_q[S_0]=m_q,\qquad
\operatorname{Var}_q(S_0)=\frac{1-m_q^2}{L}.
\end{align}
Thus, for $|m_q|<1$, the distribution of $S_0$ concentrates around $m_q$ with a standard deviation proportional to $L^{-1/2}$.

The factorized record probability also shows that $S_0$ contains all the information relevant to sector classification.
For a measurement record $\bm x$ with $N_+=j$, the conditional record probability is
\begin{align}
\Pr_q(\bm x\mid N_+=j)
=\frac{P_q^{(L)}(\bm x)}{\Pr_q(N_+=j)}
=\frac{1}{\binom Lj},
\end{align}
provided that $\Pr_q(N_+=j)>0$.
This conditional probability is independent of the sector label $q$.
Consequently, the conditional distribution of any additional record statistics, including $(S_1,\ldots,S_K)$, given $S_0$ is independent of the sector label.
The hierarchy of optimal classification fidelities in Eq.~\eqref{eq:optfid} therefore reduces to
\begin{align}
\mathcal F_{F_0}^*(L)
=\mathcal F_{F_K}^*(L)
=\mathcal F_{\rm full}^*(L),
\end{align}
for all $0\leq K<L$.
Including additional correlations or higher-order statistics cannot improve the optimal classification fidelity in this QND model.

\subsubsection{Pair-correlation statistics}

Although pair correlations provide no additional information about the sector label once $S_0$ is known, their finite-record distributions can still differ between sectors.
We first derive the mean and variance of $S_k$ at a fixed lag $1\leq k<L$.

Defining $N=L-k$ and $Z_n^{(k)}=x_nx_{n+k}$, we write $S_k=N^{-1}\sum_{n=1}^N Z_n^{(k)}$.
Since the measurement outcomes are independent, $\mathbb E_q[Z_n^{(k)}]=m_q^2$, giving $\mathbb E_q[S_k]=m_q^2$ for any record length $L>k$.
However, the products $Z_n^{(k)}$ are generally not independent.
In particular, $Z_n^{(k)}$ and $Z_{n+k}^{(k)}$ share the outcome $x_{n+k}$, giving $\mathbb E_q[Z_n^{(k)}Z_{n+k}^{(k)}]
=\mathbb E_q[x_nx_{n+2k}]=m_q^2$.
Products involving disjoint sets of outcomes are independent. 
The covariance between two products is therefore
\begin{align}
\operatorname{Cov}_q(Z_n^{(k)},Z_{n+t}^{(k)})
=
\begin{cases}
1-m_q^4,&t=0,\\
m_q^2(1-m_q^2),&t=k,\\
0,&t>0,\ t\ne k,
\end{cases}
\end{align}
where both starting positions are within the range $1,\ldots,N$.
Summing the covariance over all pairs of starting positions gives
\begin{align}
\operatorname{Var}_q(S_k)
&=\frac1{N^2}\sum_{n,n'=1}^N
\operatorname{Cov}_q(Z_n^{(k)},Z_{n'}^{(k)})
=\frac{1-m_q^4}{L-k}
+\frac{2(L-2k)_+}{(L-k)^2}m_q^2(1-m_q^2),
\label{eq:qnd_pair_variance}
\end{align}
where $(a)_+=\max(a,0)$ accounts for the absence of overlapping products when $L\leq2k$.
For fixed $k$ and large $L$, Eq.~\eqref{eq:qnd_pair_variance} gives
\begin{align}
\operatorname{Var}_q(S_k)
=\frac{1+2m_q^2-3m_q^4}{L-k}+O(L^{-2}).
\end{align}
Thus, for $|m_q|<1$, the distribution of $S_k$ concentrates around $m_q^2$ with a standard deviation proportional to $(L-k)^{-1/2}$.
In the deterministic cases $|m_q|=1$, the variance vanishes exactly.

Beyond the first two moments, we can also obtain the exact finite-record distribution of $S_k$.
Unlike the positive-outcome count $N_+$, the equal-outcome pair count $N_k$ generally does not follow a binomial distribution because overlapping products can be correlated.
To obtain its distribution, we divide the measurement record into $k$ independent subsequences,
\begin{align}
\bm x^{(a)}
=(x_a,x_{a+k},\ldots,x_{a+(\ell_a-1)k}),
\end{align}
where $a=1,\ldots,k$ and $\ell_a=1+\lfloor(L-a)/k\rfloor$ is the length of subsequence $a$.
Equal-outcome pairs at lag $k$ in the original record correspond to equal neighboring pairs within these subsequences.

To count equal neighboring pairs within a subsequence, we introduce a transfer matrix with elements $[A_q(z)]_{xy}=p_q(y)z^{\delta_{xy}}$, where $x,y\in\{+1,-1\}$ and $z$ is an auxiliary counting variable.
In the basis ordered as $(+1,-1)$, this matrix and the probability generating function for a subsequence of length $\ell$ are given by
\begin{align}
A_q(z)&=\frac12
\begin{pmatrix}
(1+m_q)z&1-m_q\\
1+m_q&(1-m_q)z
\end{pmatrix},\qquad
g_{q,\ell}(z)=\frac12(1+m_q,1-m_q)
A_q(z)^{\ell-1}
\begin{pmatrix}1\\1\end{pmatrix}.
\end{align}
Since the subsequences are independent and the total pair count $N_k$ is the sum of their equal-neighbor counts, its probability generating function factorizes as
\begin{align}
G_{q,k}^{(L)}(z)
=\mathbb E_q[z^{N_k}]
=\prod_{a=1}^k g_{q,\ell_a}(z).
\end{align}
Using $N_k=N(1+S_k)/2$, we obtain the exact finite-record distribution
\begin{align}
\Pr_q\left(S_k=\frac{2j}{N}-1\right)
=[z^j]G_{q,k}^{(L)}(z),\qquad j=0,\ldots,N,
\label{eq:qnd_pair_law}
\end{align}
where $[z^j]$ denotes the coefficient of $z^j$
in the generating function.

For unbiased outcomes ($m_q=0$), the generating function simplifies to $g_{q,\ell}(z)=[(1+z)/2]^{\ell-1}$.
Since $\sum_{a=1}^k\ell_a=L$, we obtain $G_{q,k}^{(L)}(z)=[(1+z)/2]^{L-k}$.
Consequently, $N_k$ follows a binomial distribution with parameters $L-k$ and $1/2$, giving
\begin{align}
\Pr_q\left(S_k=\frac{2j}{L-k}-1\right)
=\binom{L-k}{j}2^{-(L-k)},
\label{eq:qnd_fair_pair_law}
\end{align}
where $j=0,\ldots,L-k$.

\subsubsection{Numerical parameters and results}

In Example I, we choose $\phi=0$, $b_0T=\pi/4$, and $b_1T=\arccos\sqrt{0.8}$, giving $m_0=0$ and $m_1=-0.6$.
The two sectors have equal initial weights, $\mu_0=\mu_1=1/2$.
According to the analytical results above, the mean values of the record statistics in the two sectors are
\begin{align}
&\mathbb E_0[S_0]=0,\qquad
\mathbb E_1[S_0]=-0.6,\nonumber\\
&\mathbb E_0[S_k]=0,\qquad
\mathbb E_1[S_k]=0.36,\quad k\geq1.
\end{align}
The pair-count distribution is binomial in sector 0, while that in sector 1 is obtained from the general generating function in Eq.~\eqref{eq:qnd_pair_law}.

For each sector, we generate 10,000 measurement records of length 1000, divided into 4250 training, 750 validation, and 5000 test records.
The classification fidelities are evaluated at record lengths $L=4,8,12,20,30,50,80,120,200$, while the feature distributions are displayed at $L=1000$.
The outcome-count classifier is retained for all tested larger feature families, consistent with the analytical result that additional statistics cannot improve the optimal classification fidelity.

The joint feature distribution in the main text is constructed from the 5000 independent test records per sector at $L=1000$.
We use 99 histogram bins along each coordinate and apply Gaussian smoothing with a standard deviation of one bin for visualization.
The coordinate ranges include all displayed samples with additional margins.
The two sector-conditioned densities are normalized separately and averaged with equal sector weights.

\subsection{Example II: Two dynamical parity sectors}
\label{sec:exampleII}

\subsubsection{Parity sectors and rank-one dynamics}

The spin-pair Hamiltonian and probe coupling considered in the main text both commute with the parity operator $P=\sigma_1^z\sigma_2^z$.
In the even-parity basis $(\ket{\uparrow\uparrow},\ket{\downarrow\downarrow})$ and the odd-parity basis $(\ket{\uparrow\downarrow},\ket{\downarrow\uparrow})$, the system Hamiltonian and probe coupling reduce to two-dimensional blocks,
\begin{align}
H_q&=w_{xq}\sigma_x+w_{zq}\sigma_z,\qquad
B_q=\kappa_q\sigma_z.
\end{align}
where $q=+$ and $q=-$ denote the even- and odd-parity sectors, respectively, and the Pauli matrices act within each sector.
For each sector, the conditional evolution operators $U_{q,\pm}=e^{-i(H_q\pm B_q)T}$ can be written as
\begin{align}
U_{q,\pm}
=\cos(\omega_{q,\pm}T)I
-i\sin(\omega_{q,\pm}T)\bm n_{q,\pm}\cdot\bm\sigma,
\end{align}
where $\omega_{q,\pm}=[w_{xq}^2+(w_{zq}\pm\kappa_q)^2]^{1/2}$ and $\bm n_{q,\pm}=(w_{xq},0,w_{zq}\pm\kappa_q)/\omega_{q,\pm}$.
For the Ramsey phase $\phi=\pi/2$, the Kraus operators are given by $M_{q,x}=(U_{q,+}-ixU_{q,-})/2$.
Their determinants are
\begin{align}
\det M_{q,x}=-\frac{ix}{2}c_q,
\end{align}
where
\begin{align}
c_q
=\frac12\operatorname{Tr}(U_{q,+}^\dagger U_{q,-})
=\cos(\omega_{q,+}T)\cos(\omega_{q,-}T)
+(\bm n_{q,+}\cdot\bm n_{q,-})
\sin(\omega_{q,+}T)\sin(\omega_{q,-}T).
\end{align}
Both Kraus operators become rank one when $c_q=0$.
In the special case $w_{zq}=0$, defining $\nu_q=(w_{xq}^2+\kappa_q^2)^{1/2}$, we obtain
\begin{align}
c_q=1-\frac{2\kappa_q^2}{\nu_q^2}\sin^2(\nu_qT).
\end{align}
The rank-one condition therefore reduces to
\begin{align}
\frac{\kappa_q^2}{\nu_q^2}\sin^2(\nu_qT)=\frac12.
\label{eq:rankone_condition}
\end{align}

At rank one, each Kraus operator can be written as
\begin{align}
M_{q,x}=e^{i\chi_x}\ket{\psi_x}\bra{v_x},
\end{align}
where the sector index $q$ is omitted from the states for simplicity.
Trace preservation ensures that the two input states $\{\ket{v_x}\}_x$ form an orthonormal basis and that both Kraus operators have unit nonzero singular value.
Unitality similarly ensures that the two output states $\{\ket{\psi_x}\}_x$ form an orthonormal basis.
Therefore, each measurement prepares a state determined only by its outcome, independently of the preceding measurement history.
The probability of obtaining outcome $y$ immediately after outcome $x$ is
\begin{align}
\Pr_q(y|x)=|\langle v_y|\psi_x\rangle|^2.
\end{align}
Since the input and output states form orthonormal bases, the transition matrix is doubly stochastic and takes the symmetric form
\begin{align}
P_q=
\begin{pmatrix}
s_q&1-s_q\\
1-s_q&s_q
\end{pmatrix},
\end{align}
where $s_q=\Pr_q(x|x)$ is the probability of obtaining two identical consecutive outcomes.
For the maximally mixed initial state $\eta_q=I_2/2$, the first outcome takes either value with probability $1/2$.
Thus, the measurement record is described by a stationary symmetric first-order Markov chain.

For $w_{zq}=0$, the stay probability $s_q$ can be obtained analytically.
Since $\det M_{q,x}=0$, the Cayley--Hamilton identity gives $M_{q,x}^2=(\operatorname{Tr}M_{q,x})M_{q,x}$.
Using $\|M_{q,x}\|_F^2=1$ and $\operatorname{Tr}M_{q,x}=(1-ix)\cos(\nu_qT)$,
we obtain
\begin{align}
s_q
&=\|M_{q,x}^2\|_F^2
=|\operatorname{Tr}M_{q,x}|^2
=2\cos^2(\nu_qT).
\label{eq:rankone_stay}
\end{align}
Hence, at the rank-one condition, the statistics of the measurement record within each sector are fully determined by the corresponding stay probability $s_q$.

\subsubsection{Record statistics at rank one}

Having established the reduction to a symmetric Markov chain, we now derive the finite-record distributions of the statistical features introduced in Sec.~\ref{subsec:features}.
For each sector at rank one, the probability of a measurement record is given by the product of the first-outcome probability and the transition probabilities,
\begin{align}
P_q^{(L)}(\bm x)
=\frac12 s_q^{N_1}(1-s_q)^{L-1-N_1},
\end{align}
where $N_1=(L-1)(1+S_1)/2$ counts the number of equal adjacent outcomes.

Unlike the QND model in Example I, the individual measurement outcomes are generally not independent.
However, the adjacent products $y_n=x_nx_{n+1}$ are independent and identically distributed.
To see this, we note that the transformation from $\bm x$ to $(x_1,y_1,\ldots,y_{L-1})$ is invertible, and the corresponding probability distribution factorizes as
\begin{align}
\Pr_q(x_1,y_1,\ldots,y_{L-1})
=\frac12\prod_{n=1}^{L-1}
s_q^{(1+y_n)/2}(1-s_q)^{(1-y_n)/2}.
\end{align}
Thus, each $y_n$ takes the value $+1$ with probability $s_q$ and $-1$ with probability $1-s_q$.
Since $S_1=(L-1)^{-1}\sum_{n=1}^{L-1}y_n$, the count $N_1$ follows a binomial distribution with parameters $L-1$ and $s_q$, giving
\begin{align}
\Pr_q\left(S_1=\frac{2j}{L-1}-1\right)
=\binom{L-1}{j}s_q^j(1-s_q)^{L-1-j},
\label{eq:markov_S1_law}
\end{align}
where $j=0,\ldots,L-1$.
Defining $r_q=2s_q-1$, we obtain
\begin{align}
\mathbb E_q[S_1]=r_q,\qquad
\operatorname{Var}_q(S_1)=\frac{1-r_q^2}{L-1}.
\end{align}
For $|r_q|<1$, the distribution of $S_1$ therefore concentrates around $r_q$ with a standard deviation proportional to $(L-1)^{-1/2}$.
In contrast, stationarity gives $\mathbb E_q[S_0]=0$ in both parity sectors.
When the two sectors have different stay probabilities $s_q$ with $0<s_q<1$, their $S_1$ distributions concentrate around different values, whereas their $S_0$ distributions concentrate around the same value.

The factorized record probability also shows that $S_1$ contains all the information relevant to sector classification when both sectors are at rank one.
Following the argument in Example I, for a measurement record $\bm x$ with $N_1=j$, the conditional record probability is
\begin{align}
\Pr_q(\bm x\mid N_1=j)
=\frac{P_q^{(L)}(\bm x)}{\Pr_q(N_1=j)}
=\frac{1}{2\binom{L-1}{j}},
\end{align}
provided that $\Pr_q(N_1=j)>0$.
This conditional probability is independent of the sector label $q$.
Consequently, the hierarchy of optimal classification fidelities becomes
\begin{align}
\mathcal F_{F_0}^*(L)
\leq\mathcal F_{S_1}^*(L)
=\mathcal F_{F_K}^*(L)
=\mathcal F_{\rm full}^*(L),
\end{align}
for all $1\leq K<L$, where $\mathcal F_{S_1}^*(L)$ denotes the optimal classification fidelity using $S_1$ alone.
Thus, including additional record statistics beyond $S_1$ cannot improve the optimal classification fidelity at rank one.

We next consider the statistics of $S_k$ at larger lags.
Using the adjacent products $y_n$, the two-outcome product
at lag $k$ can be written as
\begin{align}
Z_n^{(k)}=x_nx_{n+k}=\prod_{j=n}^{n+k-1}y_j.
\end{align}
Since the variables $y_j$ are independent with mean $r_q$, we obtain $\mathbb E_q[S_k]=r_q^k$ for any record length $L>k$.
Two products $Z_n^{(k)}$ and $Z_{n+t}^{(k)}$ share $k-t$ variables when $0\leq t<k$ and are independent when $t\geq k$.
Using $y_j^2=1$, their covariance is given by
\begin{align}
\operatorname{Cov}_q(Z_n^{(k)},Z_{n+t}^{(k)})
=
\begin{cases}
1-r_q^{2k},&t=0,\\
r_q^{2t}-r_q^{2k},&1\leq t<k,\\
0,&t\geq k.
\end{cases}
\end{align}
Following the same covariance-summation procedure as in Example I, we obtain the exact finite-record variance
\begin{align}
\operatorname{Var}_q(S_k)
=\frac{1-r_q^{2k}}{N}
+\frac{2}{N^2}
\sum_{t=1}^{\min(k-1,N-1)}
(N-t)(r_q^{2t}-r_q^{2k}),
\label{eq:markov_pair_variance}
\end{align}
where $N=L-k$.
For fixed $k$ and $|r_q|<1$, the distribution of $S_k$ concentrates around $r_q^k$ with a standard deviation proportional to $(L-k)^{-1/2}$.

Beyond the first two moments, we can also obtain the exact finite-record distribution of $S_k$ using a probability generating function, as in Example I.
Since the pair count can be written as $N_k=\sum_{n=1}^{L-k}(1+\prod_{j=n}^{n+k-1}y_j)/2$, its probability generating function is
\begin{align}
G_{q,k}^{(L)}(z)
&=\mathbb E_q[z^{N_k}]
=\sum_{\bm y\in\{\pm1\}^{L-1}}
\left[
\prod_{n=1}^{L-1}
s_q^{(1+y_n)/2}(1-s_q)^{(1-y_n)/2}
\right]
z^{\sum_{n=1}^{L-k}(1+\prod_{j=n}^{n+k-1}y_j)/2}.
\label{eq:markov_pair_pgf}
\end{align}
The exact probability distribution is then given by
\begin{align}
\Pr_q\left(S_k=\frac{2j}{N}-1\right)
=[z^j]G_{q,k}^{(L)}(z),\qquad j=0,\ldots,N.
\end{align}
For $k=1$, the generating function reduces to $G_{q,1}^{(L)}(z)=[(1-s_q)+s_qz]^{L-1}$, recovering the binomial distribution derived above.
For $k\geq2$, the overlapping products generally lead to non-binomial distributions.
The generating function can be evaluated recursively by retaining the most recent $k-1$ adjacent products, without enumerating all possible measurement records.

\subsubsection{Record statistics beyond rank one}

Away from the rank-one condition, the state after each measurement generally depends on both the measurement outcome and the state before the measurement.
Consequently, the probability of the next outcome may depend on earlier measurement outcomes, and the record is not necessarily described by a first-order Markov chain.

As in the rank-one case, we initialize each sector in the maximally mixed state $\eta_q=I_2/2$, which is preserved by the unital channel $\mathcal E_q$.
The resulting measurement records are stationary.
Using the Ramsey Kraus operators, we obtain the single-outcome probability
\begin{align}
p_q(x)
&=\operatorname{Tr}(M_{q,x}\eta_q M_{q,x}^\dagger)
=\frac12(1-xc_q\cos\phi),
\end{align}
where $c_q=\operatorname{Tr}(U_{q,+}^\dagger U_{q,-})/2$ is real because $U_{q,\pm}\in SU(2)$.
Thus, at $\phi=\pi/2$, both outcomes have probability $1/2$, regardless of whether the Kraus operators are rank one or full rank.
The mean values of the record statistics are therefore
\begin{align}
\mathbb E_q[S_0]=0,\qquad
\mathbb E_q[S_k]=C_q(k),
\end{align}
where $C_q(k)=\operatorname{Tr}[ \mathcal J_q\mathcal E_q^{k-1}\mathcal J_q(\eta_q)]$ is the stationary correlation at lag $k$, and $\mathcal J_q=\sum_x x\mathcal M_{q,x}$ is the outcome-weighted map introduced in the End Matter of the main text.
Unlike the rank-one case, these correlations do not generally follow the geometric form $r_q^k$.

Although the mean of $S_0$ is zero in both sectors, its finite-record fluctuations depend on the temporal correlations.
Since $\mathbb E_q[S_0]=0$, its variance is given by $\operatorname{Var}_q(S_0)=L^{-2}\sum_{n,m=1}^L\mathbb E_q[x_nx_m]$.
Using $x_n^2=1$, stationarity, and the fact that there are $L-k$ outcome pairs at lag $k$, we obtain
\begin{align}
\operatorname{Var}_q(S_0)
=\frac1{L^2}\left[
L+2\sum_{k=1}^{L-1}(L-k)C_q(k)
\right].
\label{eq:fullrank_S0_variance}
\end{align}
Thus, although the fluctuations of $S_0$ can differ between sectors, classification based on $S_0$ alone must rely on differences in the widths or shapes of its distributions rather than on a separation of their means.
This makes sector identification difficult when these distributions strongly overlap.
In contrast, the two-outcome statistics $S_k$ can have different mean values in different sectors, providing a more effective way to distinguish them.

% The exact finite-record distributions of $S_0$ and $S_k$ can be calculated using the counting recursions described in Sec.~\ref{subsec:exact_distributions}.
% Beyond rank one, the record probability generally depends on more than the adjacent-pair count $N_1$.
% Additional correlations or higher-order statistics may therefore improve the optimal classification fidelity beyond that achievable using $F_1$.

\subsubsection{Numerical parameters and results}

To illustrate the role of temporal correlations beyond rank one, we choose model parameters such that the two sectors have nearly identical variances of $S_0$ but different mean values of $S_1$.
We set $T=1$ and $\phi=\pi/2$, with the effective block parameters
\begin{align}
(w_{x+},w_{z+},\kappa_+)
&=(1.762898273,-0.829888256,0.659490632),\nonumber\\
(w_{x-},w_{z-},\kappa_-)
&=(0.364633387,0.012000000,0.813727292).
\end{align}

The odd-parity block is constructed by slightly perturbing a rank-one point.
Defining $u=\arccos\sqrt{0.4}$, the reference parameters $w_{x-}=u/\sqrt6$, $\kappa_-=u\sqrt{5/6}$, and $w_{z-}=0$ satisfy the rank-one condition and give $s_-=0.8$.
We perturb this point by setting $w_{x-}=1.008u/\sqrt6$, $\kappa_-=1.006u\sqrt{5/6}$, and $w_{z-}=0.012$.
The even-parity block is chosen farther away from the rank-one condition.
The resulting determinant magnitudes are $|\det M_{+,x}|\approx0.36527$ and $|\det M_{-,x}|\approx0.00424$ for either outcome $x$.

Both sectors are initialized in the maximally mixed state $\eta_q=I_2/2$ and have equal initial weights $\mu_+=\mu_-=1/2$.
The first three stationary correlations are
\begin{align}
(C_+(1),C_+(2),C_+(3))
&=(0.12549412,0.13142441,0.26451114),\nonumber\\
(C_-(1),C_-(2),C_-(3))
&=(0.59153770,0.35545571,0.21351473).
\end{align}
The correlations in the odd-parity sector remain close to the geometric sequence $0.6^k$ of the rank-one reference, whereas those in the even-parity sector exhibit a different dependence on the lag $k$.
Neither sequence, however, follows an exact geometric progression.
Despite their different temporal correlations, the two sectors have nearly identical variances of $S_0$ at $L=1000$, as obtained from Eq.~\eqref{eq:fullrank_S0_variance},
\begin{align}
\operatorname{Var}_+(S_0)&=0.00395778219,\qquad
\operatorname{Var}_-(S_0)=0.00395596737.
\end{align}
Together with their identical means, this explains the strong overlap of the $S_0$ distributions observed in the main text.
In contrast, the mean values of $S_1$ differ substantially between the two sectors.

For each sector, we generate 10,000 measurement records of length 1000, divided into 4250 training, 750 validation, and 5000 test records.
We use the same record lengths and feature families as in Example I.
All classifiers are trained and evaluated using records generated from the full-rank model.
The classification infidelities of the $F_1$ classifier at $L=20,50,80,120,200$ are $0.1200$, $0.0417$, $0.0165$, $0.0056$, and $0.0004$, respectively.
At these record lengths, the $F_1$ classifier is retained for all tested larger feature budgets.
The joint feature distribution in the main text is constructed from the 5000 independent test records per sector at $L=1000$.
We use 99 histogram bins along each coordinate and apply Gaussian smoothing with a standard deviation of one bin for visualization.

\subsection{Example III: Five-spin bath}
\label{sec:exampleIII}

\subsubsection{Model and numerical parameters}

We consider a five-spin bath with the system Hamiltonian and probe coupling
\begin{align}
H_S&=\sum_{j<k}J_{jk}
\left[\sigma_j^z\sigma_k^z-
(\sigma_j^+\sigma_k^-+\mathrm{H.c.})\right], \qquad
B=\sum_jh_j\sigma_j^z.
\label{eq:dipolar_bath}
\end{align}
Both operators conserve the total excitation number $Q=\sum_j(1+\sigma_j^z)/2$, which partitions the system into six sectors $q=0,\ldots,5$ with dimensions $d_q=\binom{5}{q}$.
We initialize each sector in the maximally mixed state $\eta_q=I_{d_q}/d_q$ and assign equal weights $\mu_q=1/6$.

Unless otherwise specified, we set $T=60\,\mu\mathrm{s}$ and $\phi=\pi/2$.
The coupling strengths $J_{jk}$ and probe fields $h_j$ are listed in Table~\ref{tab:fivebath}.
With the Hamiltonian parameters expressed in kHz and $T$ in $\mu\mathrm{s}$, the conditional evolution operators are given by
\begin{align}
U_\pm=\exp[-i2\pi\times10^{-3}(H_S\pm B)T].
\end{align}

For each sector, we generate 9000 independent measurement records of length 1000, divided into 3000 training, 1000 validation, and 5000 test records.
Classification is performed at record lengths $L=20,50,100,200,500,1000$, using prefixes of the records within each data split.

\begin{table}[htbp]
\centering
\small
\begin{tabular}{crrrrr}
\toprule
$j\backslash k$ & 1 & 2 & 3 & 4 & 5\\
\midrule
1 & -- & 0.3313354983 & 0.4106532201 & -0.2846525375 & -0.5928994012\\
2 & -- & -- & -0.4190233181 & -0.3447553328 & -0.4163766368\\
3 & -- & -- & -- & -0.4980266736 & 0.0401291467\\
4 & -- & -- & -- & -- & -0.4367245289\\
5 & -- & -- & -- & -- & --\\
\midrule
$h_j$ & 1.3433304097 & 0.3772186467 & 0.3717264345 & 0.7531057964 & 0.9599170540\\
\bottomrule
\end{tabular}
\caption{Coupling strengths $J_{jk}$ and probe fields $h_j$
for the five-spin bath, in units of kHz.}
\label{tab:fivebath}
\end{table}

\subsubsection{Classification with second-order features}

We first consider the joint distribution of $S_0$ and $S_1$ shown in Fig.~4(a) of the main text.
The distribution is constructed from 5000 test records per sector at $L=1000$, using 99 histogram bins along each coordinate and Gaussian smoothing with a standard deviation of one bin for visualization.
The sector-conditioned joint distributions are combined with equal weights, while the sector-conditioned marginal distributions are normalized separately.

The complementarity of $S_0$ and $S_k$ can be understood from the spin-flip symmetry of the model.
Under the global spin flip $R=\prod_{j=1}^5\sigma_j^x$, we have $RH_SR^\dagger=H_S$ and $RBR^\dagger=-B$.
This transformation exchanges sectors $q$ and $5-q$. 
At $\phi=\pi/2$, the Kraus operators satisfy $RM_xR^\dagger=-ixM_{-x}$.
Since the maximally mixed initial states in the two sectors are also related by this transformation, their record probabilities satisfy
\begin{align}
P_q^{(L)}(\bm x)=P_{5-q}^{(L)}(-\bm x),
\label{eq:spinflip_record}
\end{align}
where $-\bm x=(-x_1,\ldots,-x_L)$.
Since $S_0(-\bm x)=-S_0(\bm x)$ and $S_k(-\bm x)=S_k(\bm x)$ for $k\geq1$, we obtain
\begin{align}
\mathbb E_q[S_0]&=-\mathbb E_{5-q}[S_0],\qquad \mathbb E_q[S_k]=\mathbb E_{5-q}[S_k],
\qquad k\geq1.
\end{align}
Thus, while the two-outcome correlations alone cannot distinguish charge-conjugate sectors, they can help distinguish sectors with overlapping $S_0$ distributions, as illustrated in Fig.~4(a) of the main text.

The classification infidelities for different feature families and the NN are listed in Table~\ref{tab:fiveerrors}.
To obtain a more reliable estimate of the small classification infidelities at $L=1000$, we generate an additional 200,000 independent test records per sector.
All classifiers are evaluated on these new records without further fitting or selection.

\begin{table}[htbp]
\centering
\small
\begin{tabular}{rrrrrrrrr}
\toprule
$L$ & $F_0$ & $F_1$ & $F_2$ & $F_7$ & $F_{16}$
& $F_{32}$ & $F_{64}$ & NN\\
\midrule
20
& 0.3959667 & 0.3693333 & 0.3693333 & 0.3693333
& 0.3693333 & -- & -- & 0.3599667\\
50
& 0.2716667 & 0.2168333 & 0.2168333 & 0.2168333
& 0.2168333 & 0.2168333 & -- & 0.2346667\\
100
& 0.1915333 & 0.1148000 & 0.1148000 & 0.1148000
& 0.1148000 & 0.1148000 & 0.1148000 & 0.1460667\\
200
& 0.1298333 & 0.0422000 & 0.0422000 & 0.0422000
& 0.0422000 & 0.0422000 & 0.0422000 & 0.0778333\\
500
& 0.0788000 & 0.0041000 & 0.0041000 & 0.0033667
& 0.0033667 & 0.0033667 & 0.0033667 & 0.0248333\\
1000
& 0.0484742 & 0.0001125 & 0.0001125 & 0.0001125
& 0.0001125 & 0.0001125 & 0.0001125 & 0.0232075\\
\bottomrule
\end{tabular}
\caption{Classification infidelities for the five-spin bath. At $L=1000$, the results are evaluated using 200,000 independent test records per sector, while at shorter record lengths, 5000 test records per sector are used. Entries with $K\geq L$ are omitted.}
\label{tab:fiveerrors}
\end{table}

Most of the improvement in classification performance comes from including the adjacent-outcome correlation $S_1$.
At $L=500$, extending the feature vector from $F_1$ to $F_7$ further reduces the infidelity from $0.004100$ to $0.003367$, while larger feature families retain the same classification rule.
At $L=1000$, all tested feature families with $K\geq1$ retain the $F_1$ classifier, which misclassifies only 135 of the 1,200,000 independent test records.
These results show that short-lag correlations provide useful classification information in this many-spin system. However, they do not establish the optimality of the selected feature-based classifiers.

\subsubsection{Convergence and finite-record fluctuations}

\begin{figure}[!htbp]
\centering
\includegraphics[width=\textwidth]{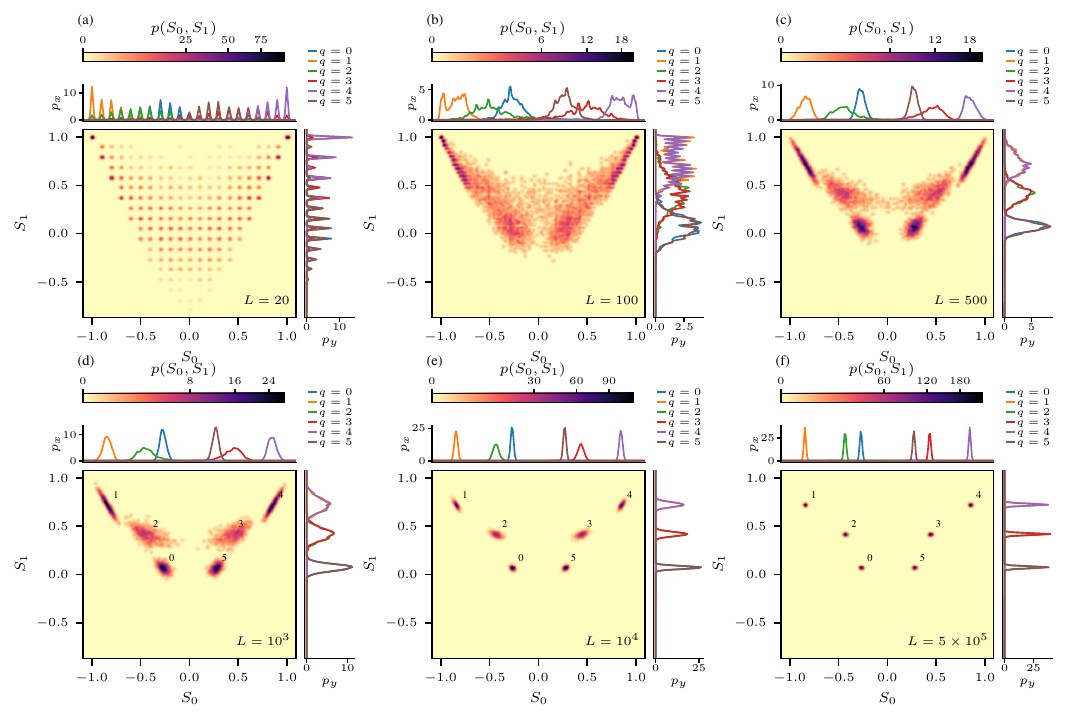}
\caption{Joint distributions of $S_0$ and $S_1$ for the five-spin bath at $T=60\,\mu\mathrm{s}$. Panels (a)--(f) correspond to $L=20,100,500,1000,10000,500000$, respectively. All panels use prefixes of the same 800 independent measurement records per sector. The coordinate ranges and histogram bins are identical across all panels, while the color scales differ. The sector-conditioned marginal distributions are normalized separately. The discrete structure at small $L$ reflects the finite set of possible feature values. }
\label{fig:fivelengthdensity}
\end{figure}

To illustrate how the record statistics converge with increasing record length, Fig.~\ref{fig:fivelengthdensity} shows the joint distributions of $S_0$ and $S_1$ at $L=20,100,500,1000,10000,500000$, using the same five-spin model as in the preceding subsection.
We generate 800 independent measurement records of length 500000 per sector and use their prefixes to obtain the statistics at different record lengths.
These records are independent of those used for classifier training, validation, and testing.
All panels use the same coordinate ranges and 199 histogram bins along each coordinate, with Gaussian smoothing of one bin for visualization.
As $L$ increases, the initially broad and overlapping distributions gradually concentrate around their sector-dependent mean values.

To quantify this concentration, we define the within-sector standard deviation $w_{q,k}(L)=\sqrt{\operatorname{Var}_q(S_k)}$ and the root-mean-square (RMS) width over all six sectors,
\begin{align}
W_k(L)=\left[\frac16\sum_{q=0}^{5}w_{q,k}^2(L)\right]^{1/2},
\qquad k=0,1.
\end{align}
This quantity characterizes the fluctuations within individual sectors and excludes the separation between their mean values. 
We estimate $w_{q,k}(L)$ from the 800 records in each sector and compare the results with the theoretical prediction derived in the End Matter of the main text,
\begin{align}
w_{q,k}^2(L)
=\frac{1}{N^2}\left[
N\gamma_{q,k}(0)
+2\sum_{t=1}^{N-1}(N-t)\gamma_{q,k}(t)
\right],
\end{align}
where $N=L-k$ and $\gamma_{q,k}(t)$ is the covariance between the samples entering $S_k$ at a separation of $t$ measurement cycles.
For $S_1$, the covariance includes the contribution from overlapping adjacent-outcome products, giving $\gamma_{q,1}(1)=C_q(2)-C_q(1)^2$.

When the covariance sum converges, defining $A_{q,k}=\gamma_{q,k}(0)+2\sum_{t=1}^{\infty}\gamma_{q,k}(t)$, we obtain the asymptotic relation
\begin{align}
\lim_{L\to\infty}\sqrt{L-k}\,W_k(L)
=\left[\frac16\sum_{q=0}^{5}A_{q,k}\right]^{1/2}.
\end{align}
For nonzero asymptotic constants, this gives the scaling relations
\begin{align}
W_0(L)\propto L^{-1/2},\qquad
W_1(L)\propto(L-1)^{-1/2}.
\end{align}

\begin{figure}[!htbp]
\centering
\includegraphics[width=\textwidth]{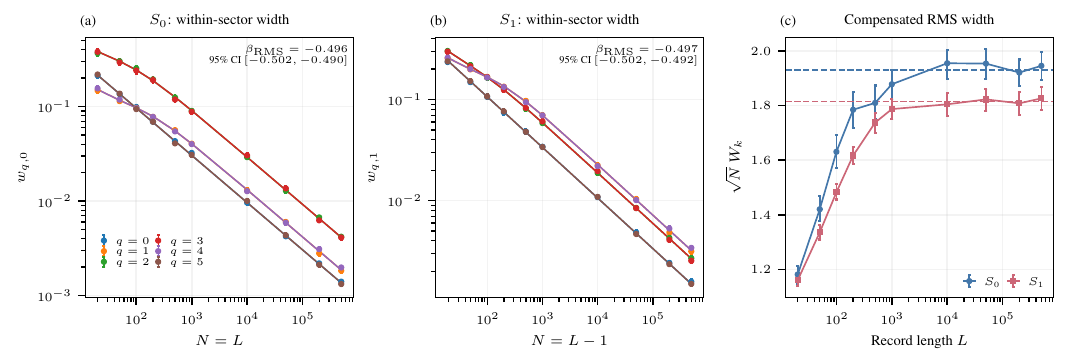}
\caption{
Finite-record fluctuations of $S_0$ and $S_1$. (a,b) Within-sector standard deviations $w_{q,0}(L)$ and $w_{q,1}(L)$, respectively. The colored points show numerical estimates with 95\% bootstrap confidence intervals, and the solid lines show the theoretical predictions. The displayed scaling exponents are obtained by fitting the RMS widths $W_k(L)$ at $L\geq1000$. (c) Compensated RMS widths $\sqrt{L}W_0(L)$ and $\sqrt{L-1}W_1(L)$. The dashed horizontal lines indicate the asymptotic values obtained from the covariance sums.
}
\label{fig:fivewidthscaling}
\end{figure}

Figure~\ref{fig:fivewidthscaling} compares the numerical estimates of the widths with the theoretical predictions. 
We evaluate the widths at $L=20,50,100,200,500,1000,10000,50000,200000,500000$.
To examine the scaling without imposing the theoretical exponent, we fit $W_k(L)\propto(L-k)^{\beta_k}$ using the five record lengths with $L\geq1000$.
The uncertainties are estimated using 1000 bootstrap resamples of the trajectories within each sector, preserving the common record prefixes across different lengths.
The fitted exponents are
\begin{align}
\beta_0&=-0.4960,\qquad
95\%~\mathrm{CI}:\ [-0.5016,-0.4899],\nonumber\\
\beta_1&=-0.4970,\qquad
95\%~\mathrm{CI}:\ [-0.5017,-0.4922].
\end{align}
Both exponents are consistent with the predicted value of $-1/2$.
Moreover, the compensated widths $\sqrt{L}W_0(L)$ and $\sqrt{L-1}W_1(L)$ approach the corresponding theoretical constants.
These results confirm both the predicted scaling and the magnitude of the finite-record fluctuations in this model.

\subsubsection{Effects of evolution time and higher-order correlations}

So far, we have considered a fixed RIM evolution time of $T=60\,\mu\mathrm{s}$.
We now investigate how the classification performance depends on $T$ and whether higher-order correlations can further improve the classification fidelity.
Keeping the Hamiltonian parameters, initial states, sector weights, and Ramsey phase unchanged, we vary the evolution time over
\begin{align}
T=30,60,90,120,160,190,230,270,320,400,500\,
\mu\mathrm{s}.
\end{align}
For each value of $T$, we use 3000 training, 1000 validation, and 5000 test records per sector, except at $T=60\,\mu\mathrm{s}$ and $L=1000$, where we use the enlarged test set described above.
The classifiers are fitted and selected separately for each combination of $T$ and $L$.

In addition to the feature vectors $F_K$, which contain correlations up to second order, we consider three extended feature vectors,
\begin{align}
&F_{64}+\bm T_3,\qquad
F_{64}+\bm T_3+U_{1,2,3},\qquad
F_{64}+\bm T_3+U_{1,2,3}+V_{2,3,4,6},
\end{align}
where $\bm T_3=(T_{1,2},T_{1,3},T_{2,3})$, and the plus sign denotes the concatenation of features.
The three feature vectors have dimensions 68, 69, and 70, respectively, and include correlations up to third, fourth, and fifth order.
Unlike the nested selection among the second-order feature families, here we retain all the specified features in the Gaussian fit and select only the covariance regularization parameter using the validation records.
Since these feature vectors contain $F_{64}$, they are evaluated only at the tested record lengths with $L>64$.

\begin{figure}[!htbp]
\centering
\includegraphics[width=\textwidth]{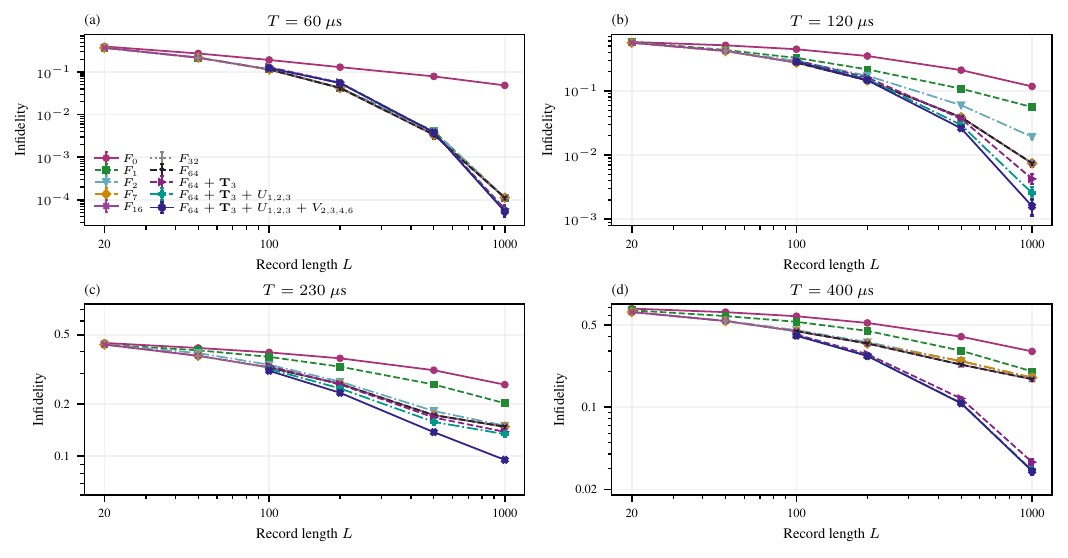}
\caption{Classification infidelities as functions of the record length $L$ for the five-spin bath at $T=60,120,230,400\,\mu\mathrm{s}$. The classifiers use different second-order feature vectors and three extended feature vectors containing higher-order correlations. The vertical ranges differ between panels. At $T=60\,\mu\mathrm{s}$ and $L=1000$, the results are evaluated using 200,000 independent test records per sector, while all other points use 5000 test records per sector. Error bars indicate approximate 95\% confidence intervals. Open downward triangles indicate one-sided 95\% upper bounds for zero observed classification errors.
}
\label{fig:fiveTlength}

\includegraphics[width=\textwidth]{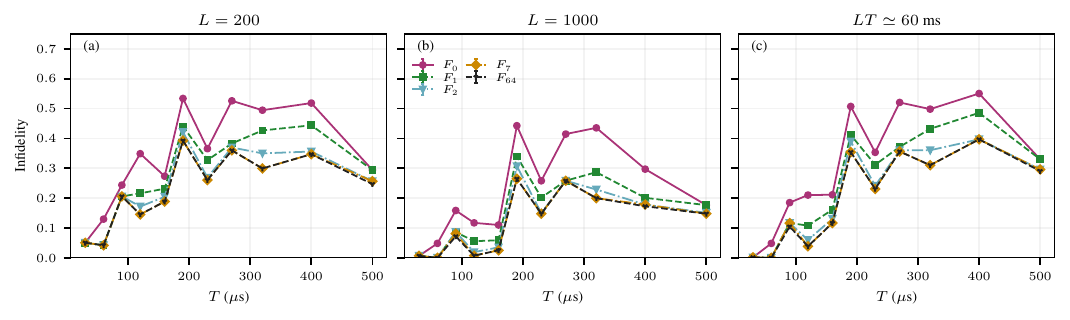}
\caption{ Classification infidelities as functions of the RIM evolution time $T$ for the five-spin bath at fixed record lengths $L=200$ and $L=1000$, and at an approximately fixed total evolution time $LT\simeq60\,\mathrm{ms}$. The feature vectors and test data are the same as those described in Fig.~\ref{fig:fiveTlength}, with the same number of test records per sector at the additional evolution times and record lengths. }
\label{fig:fiveTscan}
\end{figure}

Figure~\ref{fig:fiveTlength} compares the classification infidelities as functions of $L$ for several evolution times.
The improvement obtained by including temporal correlations depends strongly on $T$.
Including higher-order correlations does not necessarily improve the performance of classifiers fitted from finite training data.

Figure~\ref{fig:fiveTscan} further compares the classification infidelities as functions of $T$ at fixed record lengths $L=200$ and $L=1000$, as well as at an approximately fixed total evolution time $LT\simeq60\,\mathrm{ms}$, with $L=\lfloor60000/T\rfloor$ when $T$ is expressed in microseconds.
The classification performance depends nonmonotonically on $T$, because changing $T$ changes the Kraus operators and hence both the single-outcome statistics and temporal correlations of the measurement records.
For example, at $T=30\,\mu\mathrm{s}$ and $L=1000$, the $F_0$ classifier gives an infidelity of $0.006867$, and the larger second-order feature families retain this classifier.
In contrast, at $T=120\,\mu\mathrm{s}$ and the same record length, the infidelities of the $F_0$, $F_1$, and $F_7$ classifiers are $0.117300$, $0.055967$, and $0.007467$, respectively.
Thus, at some evolution times, the single-outcome statistics already provide useful classification information, whereas at other evolution times, correlations over several measurement cycles are particularly important.

In the fixed-$LT$ comparison, increasing $T$ also reduces the number of measurement cycles available within the same evolution-time budget. 
Note that $LT$ accounts only for the free evolution intervals and excludes the time required for probe rotations, initialization, and readout.

\subsection{Example IV: Six-spin bath}
\label{sec:exampleIV}

\subsubsection{Model and numerical parameters}

We consider a six-spin bath with the system Hamiltonian and probe coupling given in Eq.~\eqref{eq:dipolar_bath}.
The conserved excitation number $Q$ partitions the system into seven sectors $q=0,\ldots,6$ with dimensions $d_q=\binom{6}{q}$.
We initialize each sector in the maximally mixed state $\eta_q=I_{d_q}/d_q$ and assign equal weights $\mu_q=1/7$.
We set $T=230\,\mu\mathrm{s}$ and $\phi=\pi/2$.
The coupling strengths $J_{jk}$ and probe fields $h_j$ are independently sampled from uniform distributions,
\begin{align}
J_{jk}&\sim\mathcal U(-1.5,1.5)\,\mathrm{kHz},\qquad
h_j\sim\mathcal U(0.05,1.35)\,\mathrm{kHz}.
\end{align}
The specific realization used to illustrate the role of longer-lag correlations is listed in Table~\ref{tab:bath}.
These parameters are kept fixed throughout the simulations.

\begin{table}[htbp]
\centering
\small
\begin{tabular}{crrrrrr}
\toprule
$j\backslash k$ & 1 & 2 & 3 & 4 & 5 & 6 \\
\midrule
1 & -- & 0.76876122 & -1.30278947 & 0.16806239 & 1.23947771 & -0.63022045 \\
2 & -- & -- & -0.87151641 & -1.42165929 & 0.30585034 & 0.57930393 \\
3 & -- & -- & -- & -0.82870119 & -0.53132950 & -1.48659067 \\
4 & -- & -- & -- & -- & 1.12940112 & -0.45575381 \\
5 & -- & -- & -- & -- & -- & 0.95525585 \\
6 & -- & -- & -- & -- & -- & -- \\
\midrule
$h_j$ & 0.28824173 & 1.16710813 & 0.94447839
& 0.28660853 & 1.32159990 & 0.77323419 \\
\bottomrule
\end{tabular}
\caption{Coupling strengths $J_{jk}$ and probe fields $h_j$
for the six-spin bath, in units of kHz.}
\label{tab:bath}
\end{table}

For each sector, we generate 6000 independent measurement records of length 1000 after fixing the model parameters, divided into 3000 training, 1000 validation, and 2000 test records.
As in Example III, classification is performed at record lengths $L=20,50,100,200,500,1000$, using prefixes of the records within each data split.

\subsubsection{Classification with longer-lag correlations}

\begin{figure}[htbp]
\centering
\includegraphics[width=0.8\textwidth]{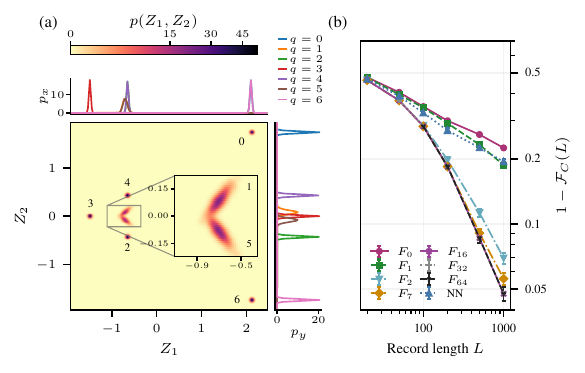}
\caption{Classification results for the six-spin bath at $T=230\,\mu\mathrm{s}$. Left: joint distribution of the first two principal component scores of $F_2$ at $L=500000$, constructed from 800 independent measurement records per sector. Right: classification infidelities of different second-order feature families and the NN as functions of the record length $L$. The PCA projection is described in the following subsection.}
\label{fig:sixspinreadout}
\end{figure}

In Example III, most of the improvement in classification performance comes from including the adjacent-outcome correlation $S_1$.
Here we show that correlations at longer lags can provide additional classification information.
Figure~\ref{fig:sixspinreadout} and Table~\ref{tab:errors} compare the classification infidelities of different feature families and the NN.

\begin{table}[htbp]
\centering
\small
\begin{tabular}{rrrrrrrrr}
\toprule
$L$ & $F_0$ & $F_1$ & $F_2$ & $F_7$ & $F_{16}$
& $F_{32}$ & $F_{64}$ & NN\\
\midrule
20   & 0.4761 & 0.4761 & 0.4616 & 0.4616
     & 0.4616 & --     & --     & 0.4674\\
50   & 0.4057 & 0.3960 & 0.3719 & 0.3719
     & 0.3719 & 0.3719 & --     & 0.3894\\
100  & 0.3486 & 0.3446 & 0.2852 & 0.2824
     & 0.2824 & 0.2824 & 0.2824 & 0.3262\\
200  & 0.2999 & 0.2919 & 0.1976 & 0.1841
     & 0.1841 & 0.1841 & 0.1841 & 0.2703\\
500  & 0.2589 & 0.2324 & 0.1125 & 0.0906
     & 0.0859 & 0.0859 & 0.0859 & 0.2250\\
1000 & 0.2245 & 0.1870 & 0.0694 & 0.0558
     & 0.0475 & 0.0475 & 0.0475 & 0.1939\\
\bottomrule
\end{tabular}
\caption{Classification infidelities for the six-spin bath, evaluated using 2000 independent test records per sector. Larger feature families may retain a classifier from a smaller family. Entries with $K\geq L$ are omitted.}
\label{tab:errors}
\end{table}

Unlike in Example III, extending the feature vector from $F_0$ to $F_1$ provides only a modest improvement, whereas including $S_2$ substantially reduces the classification infidelity.
At $L=1000$, the infidelity decreases from $0.1870$ for $F_1$ to $0.0694$ for $F_2$, and further decreases to $0.0558$ for $F_7$ and $0.0475$ for $F_{16}$.
Both $F_{32}$ and $F_{64}$ retain the $F_{16}$ classifier. 
These results show that correlations at multiple lags can improve classification, although the improvement obtained by including additional correlations depends on the record length.

The spin-flip symmetry discussed in Example III also applies to this model.
The charge-conjugate sectors $q$ and $6-q$ have opposite mean values of $S_0$ but identical distributions of the two-outcome statistics.
It is therefore important to retain $S_0$ in the feature vectors $F_K$.

To better understand how correlations at different lags help distinguish the sectors, we next examine the joint distributions of the corresponding feature vectors using principal component analysis (PCA).

\subsubsection{Principal component analysis}

To visualize the joint distributions of feature vectors with more than two components, we use principal component analysis (PCA).
For a given feature vector $F_K=(S_0,S_1,\ldots,S_K)$, we first standardize each coordinate as
\begin{align}
u_j=\frac{S_j-m_j}{d_j},
\qquad j=0,\ldots,K,
\end{align}
where $m_j$ and $d_j$ are the mean and standard deviation of $S_j$ calculated from the training records, respectively.
PCA diagonalizes the covariance matrix of the standardized feature vector $\bm u=(u_0,\ldots,u_K)$.
Let $\bm v_i^{(K)}$ denote the orthonormal eigenvectors ordered by decreasing eigenvalues.
The corresponding principal component scores are
\begin{align}
Z_i^{(K)}
&=\sum_{j=0}^K v_{ij}^{(K)}
\frac{S_j-m_j}{d_j}
=b_i^{(K)}+\sum_{j=0}^K a_{ij}^{(K)}S_j,
\label{eq:PCAaffine}
\end{align}
where $a_{ij}^{(K)}=v_{ij}^{(K)}/d_j$ and $b_i^{(K)}=-\sum_{j=0}^K a_{ij}^{(K)}m_j$.
Thus, each principal component score is an affine function of the original record statistics.

For each feature family, the standardization parameters and PCA directions are determined using the 3000 training records per sector at $L=1000$, without using the sector labels.
The standardization parameters and the coefficients in Eq.~\eqref{eq:PCAaffine} are listed in Table~\ref{tab:PCAcoefficients}.
These parameters are kept fixed when projecting independent records at different lengths.
For example, for $F_2=(S_0,S_1,S_2)$, the first two principal component scores are approximately given by
\begin{align}
Z_1^{(2)}={}&-1.319291-0.000127S_0
+1.830269S_1+1.954894S_2,\nonumber\\
Z_2^{(2)}={}&0.001293-1.825111S_0
-0.000137S_1-0.000126S_2.
\end{align}
The first principal component is dominated by the two-outcome
correlations $S_1$ and $S_2$, whereas the second is dominated
by the single-outcome statistic $S_0$.

\begin{table}[tbp]
\centering
\scriptsize
\setlength{\tabcolsep}{3pt}
\caption{
Training normalization parameters and affine coefficients
of the PCA scores.
The first two rows give the means $m_j$ and standard
deviations $d_j$ of the original statistics $S_j$.
The remaining rows give the coefficients $b_i^{(K)}$
and $a_{ij}^{(K)}$ in Eq.~\eqref{eq:PCAaffine}.
All values are rounded to six decimal places.
}
\label{tab:PCAcoefficients}
\begin{tabular}{rrrrrrrrrrr}
\toprule
& & & \multicolumn{8}{c}{Feature index $j$}\\
\cmidrule(lr){4-11}
$K$ & $i$ & $b_i^{(K)}$
& $0$ & $1$ & $2$ & $3$ & $4$ & $5$ & $6$ & $7$\\
\midrule
\multicolumn{2}{c}{$m_j$} & ---
& 0.000659 & 0.337039 & 0.359313 & 0.344320
& 0.341998 & 0.336449 & 0.331189 & 0.333735\\
\multicolumn{2}{c}{$d_j$} & ---
& 0.547912 & 0.386340 & 0.361711 & 0.371637
& 0.370780 & 0.374771 & 0.378307 & 0.375594\\
\midrule
1 & 1 & -0.616023
& -1.290549 & 1.830269 & --- & ---
& --- & --- & --- & ---\\
1 & 2 & 0.617723
& -1.290549 & -1.830269 & --- & ---
& --- & --- & --- & ---\\
\midrule
2 & 1 & -1.319291
& -0.000127 & 1.830269 & 1.954894 & ---
& --- & --- & --- & ---\\
2 & 2 & 0.001293
& -1.825111 & -0.000137 & -0.000126 & ---
& --- & --- & --- & ---\\
\midrule
7 & 1 & -2.410006
& 0.000125 & 0.970332 & 1.041595 & 1.019044
& 1.021878 & 1.011167 & 1.001728 & 1.007933\\
7 & 2 & 0.001020
& -1.825111 & 0.000586 & 0.000639 & -0.000098
& 0.000056 & -0.000469 & -0.000156 & -0.000062\\
\bottomrule
\end{tabular}
\end{table}
\subsubsection{Projected feature distributions}

\begin{figure}[tbp]
\centering
\includegraphics[width=\textwidth]{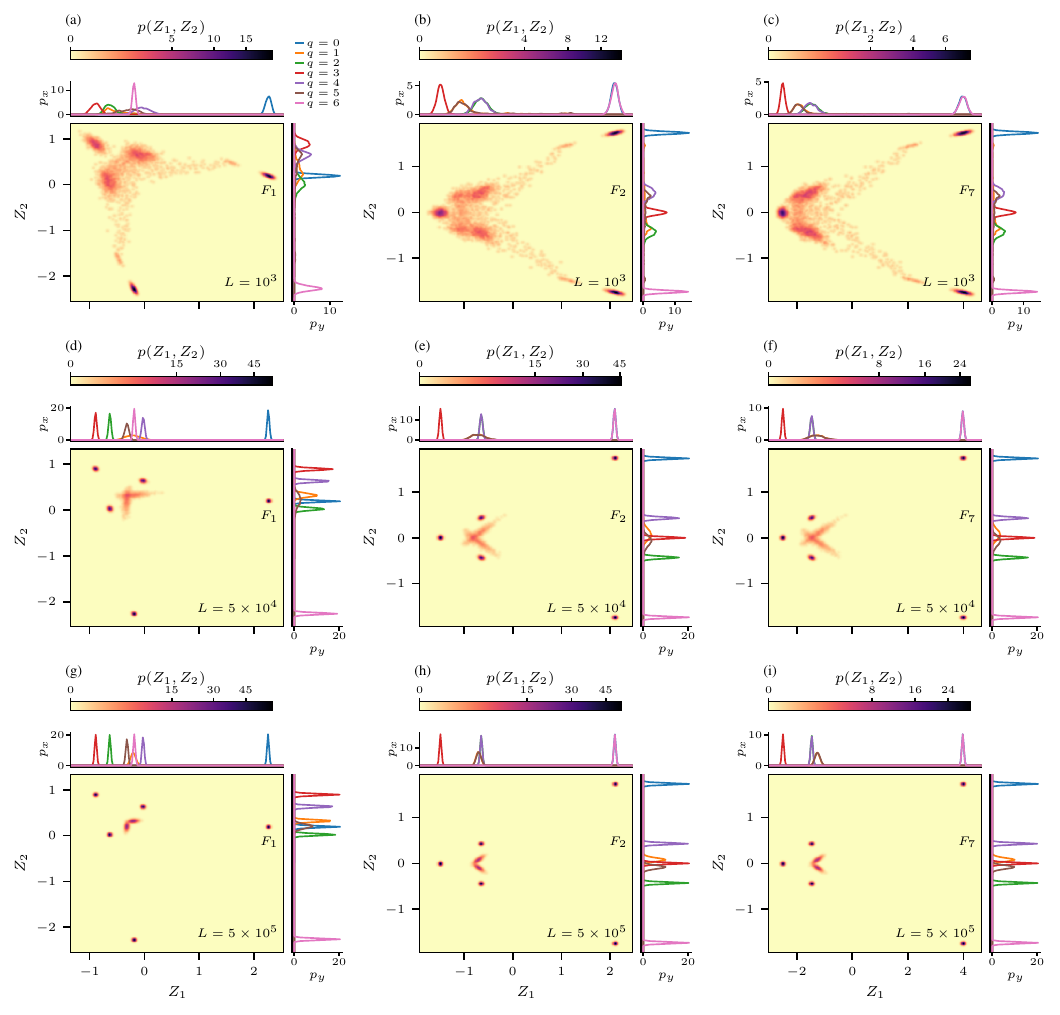}
\caption{Projected joint feature distributions for the six-spin bath. The three columns correspond to $F_1$, $F_2$, and $F_7$, while the three rows correspond to $L=1000,50000,500000$, respectively. All panels use prefixes of the same 800 independent measurement records per sector. Within each column, the standardization parameters, PCA directions, coordinate ranges, and histogram bins are kept fixed across record lengths. Each panel has its own density scale, and the sector-conditioned marginal distributions are normalized separately. The sector-color legend in panel (a) applies to all panels.}
\label{fig:PCAfamilies}
\end{figure}

Using the PCA projections defined above, we now examine the joint feature distributions for different maximum correlation lags $K$ and record lengths $L$.
For each sector, we generate 800 independent measurement records of length 500000 and use their prefixes to evaluate the feature vectors at different record lengths.
Figure~\ref{fig:PCAfamilies} compares the joint distributions of the first two principal component for $F_1$, $F_2$, and $F_7$ at $L=1000,50000,500000$.
As in Example III, the feature distributions gradually concentrate around their sector-dependent centers as $L$ increases.

\begin{figure}[!ht]
\centering
\includegraphics[width=0.88\textwidth]{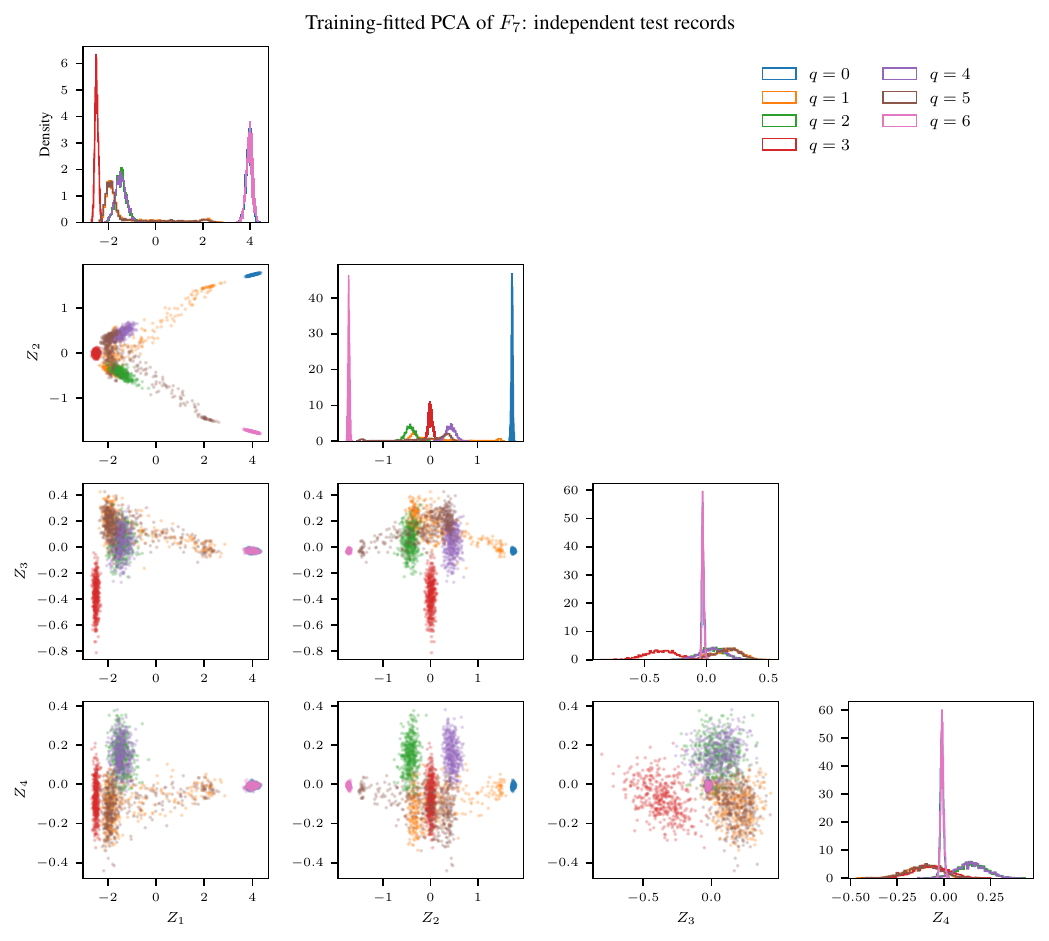}
\caption{Marginal distributions and pairwise projections of the first four principal component scores of $F_7$. The diagonal panels show the sector-conditioned marginal distributions, normalized separately. The lower panels show scatter plots of every fifth independent test record. Different colors represent different excitation sectors $q=0,\ldots,6$. The standardization parameters and PCA directions are determined using only the training records.}
\label{fig:F7projections}
\end{figure}

Comparing different feature families at the same record length illustrates how including additional correlations changes the joint feature distributions.
To illustrate this, Fig.~\ref{fig:F7projections} shows the pairwise joint distributions and marginal distributions of the first four principal component scores of $F_7$.
Sectors that overlap in one projection can exhibit different distributions in another.
Thus, different projection directions can reveal complementary information contained in the original feature vector.

\bibliography{main}